\documentclass{nature}
\usepackage{eqnarray}
\usepackage[utf8]{inputenc}
\usepackage[T1]{fontenc}
\usepackage[english]{babel} 
\usepackage[]{layout}
\usepackage{amsmath,amsfonts,amssymb}

\usepackage{graphicx}
\usepackage{float}
\restylefloat{table}
\usepackage{bm}
\usepackage{color}
\usepackage{braket}
\usepackage{array,multirow,makecell,multicol}
\usepackage[markup=default]{changes}
\usepackage{multirow}
\makeatletter
\let\saved@includegraphics\includegraphics
\AtBeginDocument{\let\includegraphics\saved@includegraphics}
\renewenvironment*{figure}{\@float{figure}}{\end@float}
\makeatother
\usepackage{diagbox}
\usepackage{upgreek} 

\title{Observation of a supersolid phase in a spin-orbit coupled exciton-polariton Bose-Einstein condensate at room temperature}

\author{Marcin\,Muszy\'nski$^{1\dagger}$, Pavel\,Kokhanchik$^{2\dagger}$, Rafa\l{}\,Mirek$^{3\dagger}$, Darius\,Urbonas$^3$, Pietro Tassan$^3$,
Piotr\,Kapu\'sci\'nski$^1$, Przemys\l{}aw\, Oliwa$^1$,  Ioannis Georgakilas$^3$,
Thilo St\"{o}ferle$^3$, Rainer F. Mahrt$^3$, Michael Forster$^4$, Ullrich Scherf$^4$,
Dmitriy~Dovzhenko$^5$, Rafa\l{}\,Mazur$^6$, Przemys\l{}aw\,Morawiak$^6$, Wiktor\,Piecek$^6$,
Przemys\l{}aw\,Kula$^7$, Barbara\,Pi\k{e}tka$^1$, Dmitry\,Solnyshkov$^{2,8}$,    Guillaume\,Malpuech$^{2\ast}$, Jacek\,Szczytko$^{1\ast}$}

\begin{document}
\maketitle

\begin{affiliations}
\item Institute of Experimental Physics, Faculty of Physics, University of Warsaw, Poland
\item Universit\'e Clermont Auvergne, Clermont Auvergne INP, CNRS, Institut Pascal, F-63000 Clermont-Ferrand, France
\item IBM Research Europe -- Zurich, S\"{a}umerstrasse 4, 8803 R\"{u}schlikon, Switzerland
\item Macromolecular Chemistry Group and Wuppertal Center for Smart Materials \& Systems (CM@S), Bergische Universit\"{a}t, Wuppertal, Gauss Strasse 20, 42119 Wuppertal, Germany
\item School of Physics and Astronomy, University of Southampton, Southampton SO17 1BJ, United Kingdom
\item Institute of Applied Physics, Military University of Technology, Warsaw, Poland
\item Institute of Chemistry, Military University of Technology, Warsaw, Poland
\item Institut Universitaire de France (IUF), 75231 Paris, France
\item[$\dagger$] These authors have contributed equally to the work.
\end{affiliations}

\begin{abstract}
In Bose-Einstein condensates (BEC), spin-orbit coupling (SOC) produces supersolidity. It is a peculiar state of matter, which, in addition to the superfluid behaviour shows periodic density modulation typical for crystals. Here, we report the fabrication of a new type of optical microcavity allowing to achieve room-temperature supersolidity for a quantum fluid of light. The microcavity is filled with a nematic liquid crystal (LC) and two layers of the organic polymer MeLPPP hosting exciton resonances. We demonstrate exciton-polariton condensation in the two distinct degenerate minima of the dispersion created by the LC induced Rashba-Dresselhaus (RD) SOC. The condensate real-space distribution shows density stripes located randomly from one condensate realization to another despite the presence of a random disorder potential. This demonstrates the immunity of stripes against disorder (that is, superfluidity) and the spontaneous breaking of translational invariance. We also report the random appearance of vortices via the Kibble-Zurek mechanism, another smoking gun of superfluidity.
\end{abstract}

\maketitle

\begin{figure}
\centering
\includegraphics[width=0.9\linewidth]{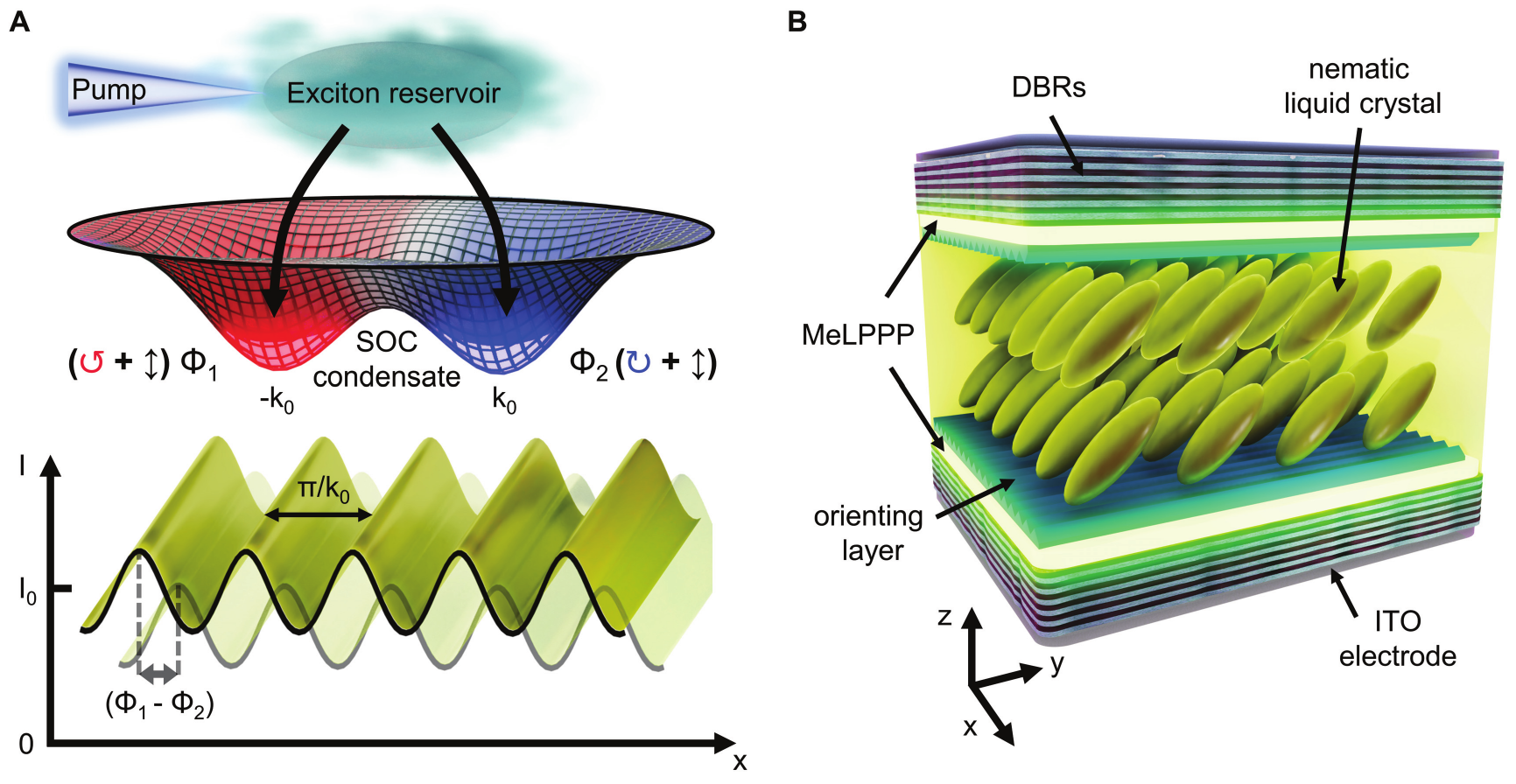}
\caption{\label{fig:Fig_0} \textbf{SOC condensate and microcavity scheme} 
(a) SOC condensate forms with different elliptical polarizations and phases $\Phi_1$ and $\Phi_2$ in the two dispersion minima. Their interference results in the formation of density stripes in real space with position depending on $\Phi_1-\Phi_2$ (b) Microcavity scheme: thick LC layer sandwiched between two organic polymer layers (MeLPPP), two DBRs, and two ITO electrodes. }
\end{figure}

Supersolidity is an intriguing state of matter that combines two antagonistic features, namely solid and superfluid properties. Here, by solid, we mean crystalline order or periodic modulation of a medium. It is combined with a frictionless flow in the presence of rugosity. 
Supersolidity was first conjectured in helium \cite{leggett1970can}, but was found difficult to observe \cite{chan2013overview}. The supersolid phase is expected to play a key role in the behavior of neutron stars being a possible explanation for the abrupt changes of their rotation frequency (glitches)~\cite{Poli2023}. 
Condensates of weakly interacting bosons exhibit superfluidity, and different methods have been implemented to break their translational invariance, hoping to preserve at the same time their superfluid properties. The first way was to consider the action of optical lattices, which can even be continuously tuned \cite{leonard2017supersolid}, with the drawback that the spatial symmetry is not broken spontaneously. The second way was to engineer a so-called Rashba-Dresselhaus spin-orbit coupling \cite{koralek2009emergence,lin2011spin} in a spinor BEC \cite{li2017stripe,geier2023dynamics}. As sketched in Fig.~\ref{fig:Fig_0}(a), the RDSOC deeply modifies the mode dispersion leading to the formation of two degenerate energy minima with opposite non-zero wavevectors $k_0$. Condensates which form in these two minima interfere, creating density stripes in real space. The wavevector $k_0$ governs the period, whereas the position of the stripes depends on the two spontaneously broken phases $\Phi_1$ and $\Phi_2$. The third way to create a supersolid is to use atoms with dipolar interactions that form periodic droplet patterns, for which superfluidity \cite{tanzi2021evidence,biagioni2024measurement} and even vortices have recently been reported \cite{casotti2024observation}. Supersolidity has also been suggested in other systems, such as hybrid Bose-Fermi mixtures~\cite{Matus2012}. All these experiments were done with ultracold atomic gases at sub-microkelvin temperatures.

In addition to cold atoms, photonic platforms, such as microcavities and photonic crystals, also host macroscopically populated states of interacting bosons (exciton-polaritons \cite{carusotto2013quantum,kavokin2017microcavities}) for which superfluidity was demonstrated both under resonant \cite{amo2009superfluidity,lerario2017room} and non-resonant excitation \cite{caputo2018topological}. 
Density stripes have also been recently reported \cite{trypogeorgos2025emerging} in a polaritonic crystal slab at cryogenic temperature hosting bound states in the continuum with negative effective mass, in which non-superfluid bright soliton condensates are formed \cite{ardizzone2022polariton,riminucci2023polariton,septembre2024soliton}. The polaritonic modes were organized in such a way to allow the condensate depletion by resonant parametric scattering. As a result, the soliton condensate interferes with the signal and idler being at the same energy, which leads to a spatially modulated amplitude of the wave function. This has been followed up by another recent report of a polariton stripe condensate~\cite{zhai2025electricallytunablenonrigidmoire}.

Photonic platforms also represent paradigmatic systems to engineer 2D photonic modes demonstrating topological singularities and topological transitions~\cite{kavokin2005,leyder2007,terccas2014non,nalitov2015polariton,klembt2018exciton,richter2019voigt,Rechcinska_Science2019,gianfrate2020measurement,ren2021nontrivial,krol2021observation,polimeno2021tuning,spencer2021spin,long2022helical,krol2022annihilation,liang2024polariton}.
The key ingredient behind this richness are the different types of effective SOC of light, but also the full tomographic access to the particle wavefunctions \cite{gianfrate2020measurement}. The most prominent SOC, which was described quasi-simultaneously in microcavities~\cite{kavokin2005} and in other systems~\cite{bliokh2008geometrodynamics}, is related to the transverse (TE-TM) nature of light modes. Another one is an emergent RDSOC, which was recently implemented in microcavities~\cite{Rechcinska_Science2019}. It occurs when Fabry-P\'erot modes of different parities and different polarizations are brought in resonance by a large linear birefringence and when the optical axis of the media filling the cavity differs from the one of the cavity itself. In that perspective, the realization of cavities filled with nematic LCs~\cite{Rechcinska_Science2019,krol2021observation,lempicka2022electrically,li2022manipulating}  represented a breakthrough, since it allows to tune linear birefringence and to control RDSOC on demand. A key aspect of RDSOC with respect to TE-TM SOC is that it scales linearly with the wavevector, and therefore, the resulting dispersion shows two degenerate minima at $\pm k_0$. 
The cavity exciton-polariton platform is therefore ideal for the study of the behaviour of quantum fluids \cite{carusotto2013quantum} in the presence of SOC and in topologically non-trivial bands.

The combination of strong exciton-photon coupling with mode tunability including the RDSOC regime has so far been achieved using lead halide perovskites as an active material~\cite{lempicka2022electrically,LempickaMirekNanophotonics,li2022manipulating,liang2024polariton}. These materials suffer from limited stability against photo-excitation.
As an alternative, organic polymer MeLPPP~\cite{scherf1992melppp} appears quite suitable, as it has demonstrated exciton-polariton condensation at room temperature~\cite{plumhof2014room,zasedatelev2021single}. Its high saturation density, quantum efficiency and photostability allow single shot measurements, which can be repeated a large number of times at a given sample location.

In this work, we report the fabrication and optical characterization of original microcavities with embedded nematic LC and MeLPPP (scheme shown in Fig.~\ref{fig:Fig_0}(b)). We prove the formation of strongly coupled exciton-photon modes with Rabi splittings of the order of 90~meV. We demonstrate the tunability of these modes in a wide spectral range by an external electric field controlling the linear birefringence, thereby reaching the RDSOC regime in which polariton condensates demonstrate density stripes. In single shot experiments and for sufficiently large condensate densities, we observe the stripe patterns placing themselves randomly from one condensation process to another, in agreement with theoretical simulations. This demonstrates the spontaneous breaking of translational invariance. The preservation of spontaneous symmetry breaking despite the presence of the disorder potential of the samples is a typical feature of a superfluid. For smaller condensate densities (close to threshold), we find the stripes to be partly pinned by disorder. We also observe the presence of quantized vortices in about 10\% of the condensate realizations.

\begin{figure}
\centering
\includegraphics[width=1\linewidth]{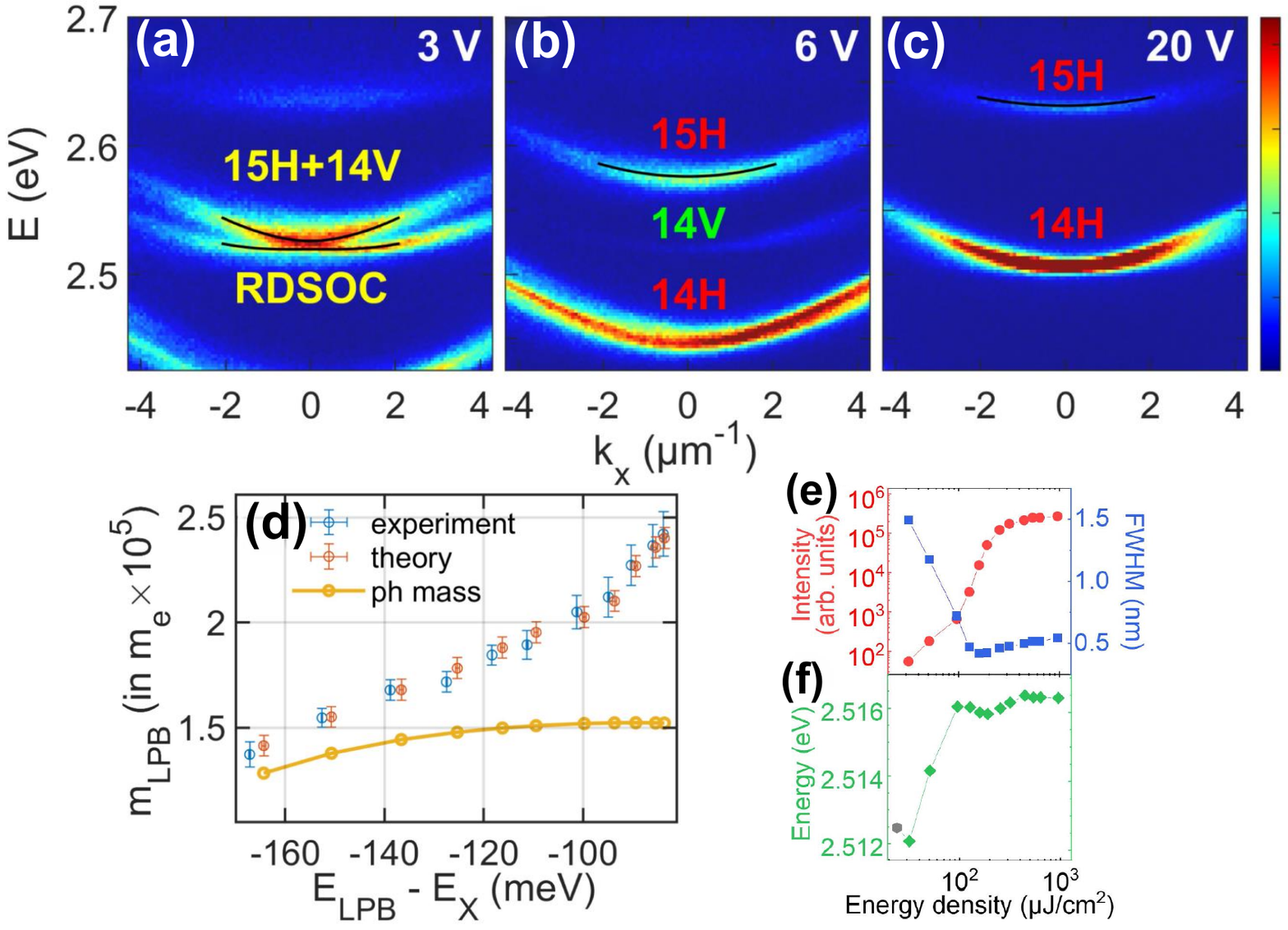}
\caption{\label{fig:Fig_1} \textbf{Demonstration of strong coupling regime.} 
 (a,b,c) experimentally measured angle-resolved PL for different values of voltage, showing tunability of the H polarized modes and (a) reach of the RDSOC regime; red, green and yellow text labels the number and polarization of the mode(s); black solid lines show the fitting (a) by Hamiltonian~\eqref{RDSOC} and (b,c) by parabola; the excitonic resonance is centered at $E_X=2.715$~eV; (d) LPB mass as a function of LPB detuning: extracted from the experiment (blue dots) and numerical simulation (red dots); numerical model uses the mode photonic mass at each voltage (yellow dots);(e) Emission intensity (red dots), linewidth (blue squares), and (f) mode energy (green diamonds) as a function of the excitation density. An extra point from reflectivity (gray hexagon) added for comparison. }
\end{figure}

\section*{Microcavity fabrication and characterization}
A scheme of the fabricated samples is shown in Fig.~\ref{fig:Fig_0}(b) (sample preparation details are given in Methods). A typical sample consists of SiO$_2$/Ta$_2$O$_5$ distributed Bragg reflectors (DBRs), which form a Fabry-P\'erot resonator hosting a series of longitudinal modes labeled by their integer order. A nematic LC is embedded within the cavity. When a voltage is applied to the ITO electrodes, the LC molecules rotate in the $yz$ plane defined by orienting layers, thus providing a tunable birefringence. The LC refractive indices along the ordinary and extraordinary axes are $n_o\approx1.57$ and $n_e\approx1.98$, respectively. Modes polarized along $y$ under normal incidence (H) are tunable by LC rotation, while modes polarized along $x$ (V) are not affected by voltage. However, both can be tuned by changing the position on the sample thanks to a slightly wedged cavity.
Two 35 nm-thick layers of methyl-substituted ladder-type poly(p-phenylene) (MeLPPP; $M_n = 31,500$, $M_w = 79,000$) surround the LC layer. Details on the optical response of MeLPPP excitons are provided in Methods (Ext. Data Fig.~\ref{fig:sm_1}) and in Refs.~\cite{plumhof2014room,zasedatelev2019room}. We fabricated a series of similar samples, because the optical quality of the MeLPPP, in spite of its high photostability, was found to degrade in time, probably due to interplay with LC molecules, with a scale ranging from a few days to a few weeks.

The samples were analyzed by angle-resolved photoluminescence (PL) (see Methods). Fig.~\ref{fig:Fig_1}(a-c) shows for the sample "1", several of these spectra detected in H-polarization. The modes 15H and 14V enter in resonance at 3~V (Fig.~\ref{fig:Fig_1}(a)) giving rise to the typical "Rashba" spectrum (RDSOC regime). At all voltages, the dispersion and polarization at small $k_x$ are described (black lines in Fig.~\ref{fig:Fig_1}(a-c)) by a 2x2 Hamiltonian (written on the circular basis) including the RDSOC coupling between H and V~\cite{Rechcinska_Science2019} which reads:
\begin{equation}
    H_{RDSOC}=\frac{\hbar^2k^2}{2M}I+(\Delta+\gamma k^2)\sigma_x+2\alpha k_x \sigma_z
 \label{RDSOC}
\end{equation}
with $I$ the identity matrix, $\sigma_{x,z}$ are Pauli matrices, $\Delta=E_V-E_H$ is the H-V detuning (corresponding to Raman coupling in atomic systems), $M=2m_Vm_H/(m_H+m_V)$, $\gamma=\hbar^2/2m$, $m=2m_Vm_H/(m_H-m_V)$, $\alpha$ the RDSOC coupling constant. 

With increasing voltage (Fig.~\ref{fig:Fig_1}(a-c)), the mode 15H quits the strong RDSOC regime and approaches the exciton resonance energy $E_X=2.715$~eV from below. We see in this figure and all other samples as well, that the mass of the modes as they approach the exciton is strongly increasing, indicating strong mixing of the photon with the exciton resonance. The mass of this lower polariton branch (LPB) is extracted by fitting the PL with a parabola, shown by the black solid line. The mass dependence versus LPB detuning $E_{LPB}-E_X$ (controlled by voltage) is displayed in Fig.~\ref{fig:Fig_1}(d). In order to fit this experimental dependence and extract the value of Rabi splitting $\Omega$ we use a Tavis-Cummings model (see Methods) taking into account the inhomogeneous broadening for excitons in MeLPPP. The model includes a single photonic mode, defined by its mass and energy obtained from the Berreman method simulation (see  Methods, "Photonic Mass"), and $N$ excitons, having a Gaussian distribution of their energies and each coupled to a photonic mode with the coupling strength $\hbar g=\Omega/\sqrt{N}$. Therefore, for each voltage, our theoretical model provides both the LPB energy and mass which we plot in Fig.~\ref{fig:Fig_1}(d). The extracted Rabi splitting is $\Omega=93$~meV. The constructed model is equivalent to the coupled oscillator model with two oscillators coupled by $\Omega$ for large exciton-photon detunings, while the behavior of the two models diverges when the detuning is reduced.
We also extracted the LPB exciton fraction of $\approx5.5\%$ in the RDSOC regime (Fig.~\ref{fig:Fig_1}(a)). These results confirm the achievement of exciton-photon strong coupling in the presence of RDSOC at negative exciton-photon detuning.  This feature is robust and qualitatively repeated for all samples studied in this work. At low detunings, an anti-crossing can be observed in a part of the samples (see Methods, "Anticrossing"), but not always, because of the cumulative inhomogeneous and temperature-induced broadening comparable with the Rabi splitting $\Omega$. Nevertheless, in all samples, the modes are strongly coupled exciton-photon states at negative exciton-photon detuning (see Methods).

\section*{Observation of stripe phase}

We now consider strong optical pumping for which room temperature exciton-polariton condensation occurs. We use pulsed non-resonant excitation (see Methods, "Experimental details"), and our observations are based both on multiple-shot experiments averaging over 500-2500 pulses and on single-shot ones, for which only measurements above the threshold provided a sufficient signal. First, we consider well-separated linearly-polarized modes. Fig.~\ref{fig:Fig_1}(e,f) shows, for one set of parameters, the PL emission intensity, linewidth, and energy of the V-polarized polariton mode versus pumping density. The exponential increase in emission intensity, the narrowing of the linewidth, and the weak blueshift provide evidence of polariton condensation. The threshold density of 95~$\mu$J/cm$^2$ is comparable to previous reports of MeLPPP polariton condensates~\cite{plumhof2014room,zasedatelev2019room,scafirimuto2018melppptunable}.
Polariton condensation is assisted by a vibron resonance and occurs about 200~meV below the excitonic peak in modes with 5$\%$ exciton fraction. The blueshift of 4~meV  is almost 3 times lower than the 11~meV difference between the LPB and bare photon energies (exciton-photon detuning of $-190$~meV). This demonstrates that the strong coupling regime holds above the condensation threshold. Also, spatial coherence measurements \textcolor{black}{in the parabolic regime}  are shown in Ext. Data Fig.~\ref{fig:coher} (Methods). Again, these features are robust from one sample to another.

\begin{figure}
\centering
\includegraphics[width=0.65\linewidth]{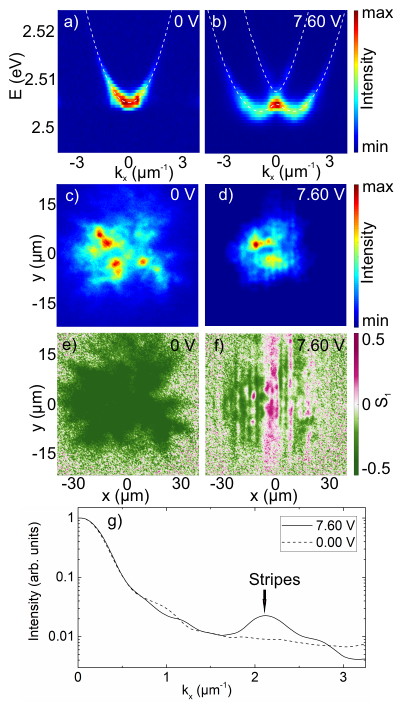}
\caption{\small \label{fig:Fig_3} \textbf{Stripe phase in a polariton condensate.} Photoluminescence measured at 0~V (first column), 7.6~V in the presence of RDSOC (second column). The first row (a,b) shows the $k_x$-dependent total emission with the double minima dispersion in (b). The second row (c,d) shows the real space total emission with stripes absent in (c) and present in (d). The third (e,f) rows show the real space distribution of the linear $S_1$ polarization degree. (g) Fourier transform of the experimental intensity from (c,d) with a marked stripes diffraction peak.}
\end{figure}

In the next part, we study the condensation in the RDSOC regime, first in multiple-shot experiments (500 shots). 
Fig.~\ref{fig:Fig_3}(a) shows the k-space emission just above the condensation threshold at 0~V, when linearly-polarized modes are well separated, while the corresponding real-space emission is shown in Fig.~\ref{fig:Fig_3}(c). It is affected by disorder, showing bright spots on a homogeneous background of 25~$\mu$m diameter, comparable with the pump spot size. \textcolor{black}{The characteristic amplitude of the disorder potential acting on polaritons is given by the inhomogeneous linewidth of the mode emission below the threshold which is a few meV. Fluctuations are of micrometer size. A microscopic image of the sample together with the spatially resolved energy spectrum is shown in Fig.~S7 in Supplementary Materials}. The emission is fully linearly polarized (Fig.~\ref{fig:Fig_3}(e)). The Fourier transform of the real space density (Fig.~\ref{fig:Fig_3}(g), dashed line) demonstrates the absence of any spatial periodicity.

By changing voltage (on the same sample point), two modes of opposite parity come into resonance, turning on the RDSOC (Fig.~\ref{fig:Fig_3}(b)). The bands show two minima at $\pm k_0\approx 1$~$\mu$m$^{-1}$. The dispersion is well described by the effective Hamiltonian~\eqref{RDSOC} with $\Delta=-1.3$ meV and $\alpha=2.7$ meV$\cdot \mu$m. 
This Hamiltonian yields $k_0$ as: 
\begin{equation}
    k_0\approx \frac{2M}{ \hbar^2}\sqrt{{\alpha^2}-\frac{\hbar^2\Delta^2}{4\alpha^2}}.
\end{equation}
The eigenstates at the two minima are elliptically-polarized with opposite circular polarization degree and same linear polarization degree (see  Methods, "Analytical description..."). The experimentally-measured linear polarization degree is $16\%$. Its analytical value is:
\begin{equation}
    S_1=\frac{\Delta+\gamma k_0^2}{\sqrt{4\alpha^2 k_0^2+(\Delta+\gamma k_0^2)^2}},
\end{equation}
yielding 23\%. If one considers an equilibrium BEC forming in these two degenerate dispersion minima, the interference between these components generates polarization stripes (spin helix), whose contrast is given by the circular polarization degree ($\approx 80\%$), and density stripes with a contrast given by the linear polarization degree ($\approx 16\%$). The stripes period is theoretically given by $\pi/k_0$. If the stripes are located completely randomly from one experiment to another, the contrast of an averaging over N realizations should be reduced by $1/\sqrt{N}$ (See Section~4 in SM) which would give 0.7\%.


The actual real space emission (Fig.~\ref{fig:Fig_3}(d)), besides being affected by disorder (as at 0~V), indeed shows vertical stripes parallel to the $y$-axis. The regularity of this modulation is demonstrated in Fig.~\ref{fig:Fig_3}(g) where the Fourier transform of the real space emission (solid line) demonstrates a peak at $k_x=2k_0 \approx 2.1~\mu$m$^{-1}$, in reasonable agreement with the one expected from direct k-space measurements. Please note that the symmetric part of the Fourier transform for $k<0$ is not shown. Also, as explained in more details in the next section, the spatial extension of stripes sets a minimal value for the coherence length of the condensate wavefunction. 
Section~1 in SM shows stripes measured at another point with a slightly different voltage.
The ratio between the intensity of the diffraction peak and the one at $k_x=0$  gives an estimate of the stripes contrast ($2.2\%$) which is three times more than the expected theoretical value, demonstrating the occurrence of a moderate pinning mechanism, either by the pump itself or by the disorder potential. The polarization stripes period (Fig.~\ref{fig:Fig_3}(f)) is similar to the one of the density stripes, whereas their contrast is $40\%$.

\section*{Spontaneous breaking of translational invariance, spatial coherence and superfluidity}

\begin{figure}
\centering
\includegraphics[width=0.73\linewidth]{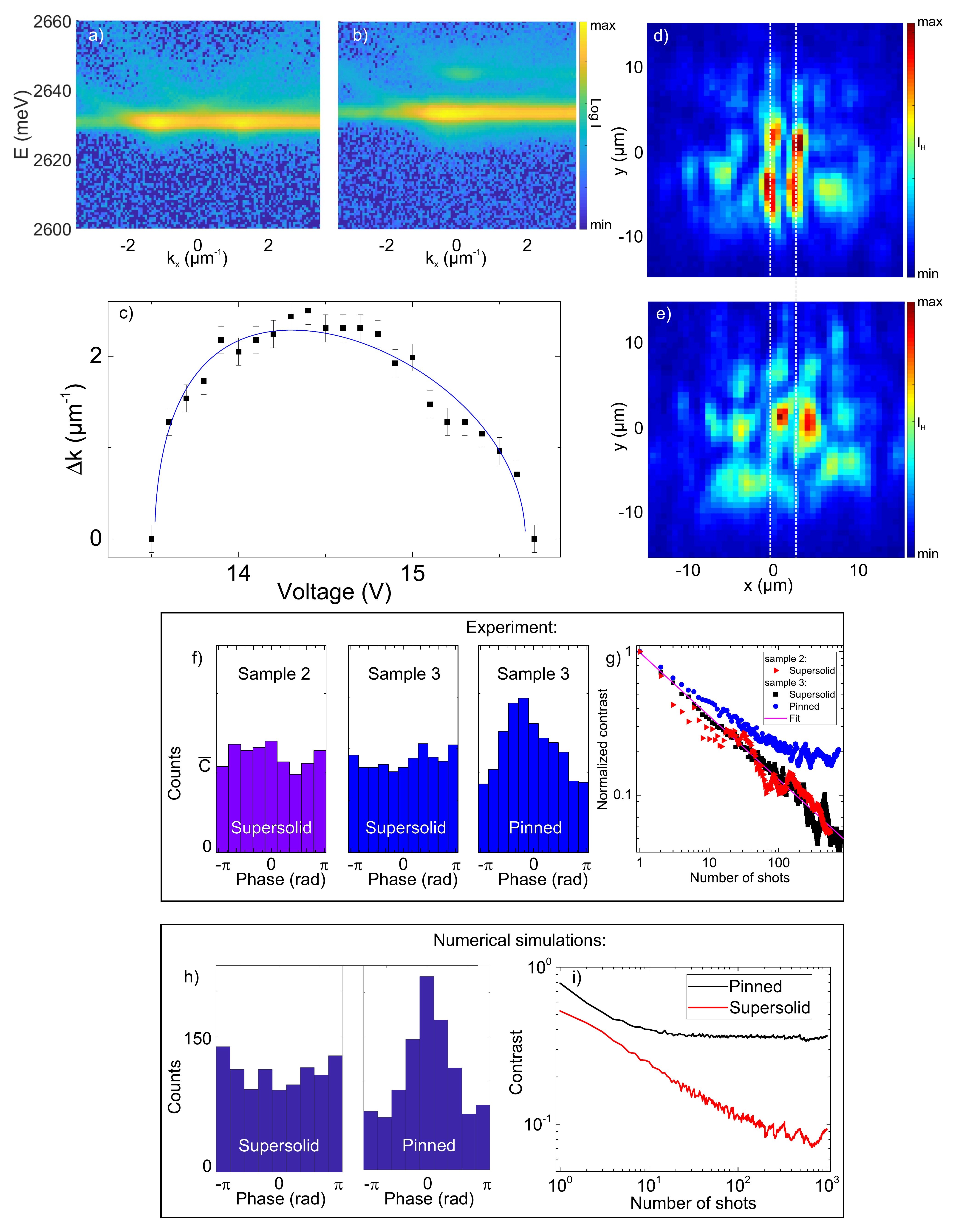}
\caption{\footnotesize \label{fig:Fig_4} \textbf{Spontaneous symmetry breaking and superfluidity evidence in single-shot experiments and simulations.} a),b) k-space saturated emission at 14.5 and 13.5 V respectively. c) k-space spacing between the emission maxima versus voltage. The blue line is the theory (See Section~2 in SM).  d), e) Two single-shot images of the condensate emission with a linear polarizer in front of the camera. White dashed lines are the guides for the eyes allowing to visualize the shift of the fringes due to the spontaneous symmetry breaking (random phase).  f) The histogram of the values of the phase of the fringes pattern. The two left histogram are taken for high condensate density and show a near-uniform distribution. The right panel is taken at lower density and shows a peak. g) The decrease of the contrast due to averaging for integrated multiple-shot experiments. Red corresponds to the left panel. Black corresponds to the central panel, blue to the right panel. The contrast decays as $1/\sqrt{N}$ at large $N$. (\textbf{h,i}) Simulations. h) Phase histogram for interacting (left) and non-interacting condensate (right) Interaction energy is 1 meV and disorder amplitude 0.5 meV. (\textbf{i}) Contrast averaged versus  number of shots with interactions (red) and without (black).}
\end{figure}

We next present in Fig.~\ref{fig:Fig_4} the results of measurements on sample "2" and sample "3", similar to sample "1" \textcolor{black}{and all being in the strong coupling regime}. The experiments performed were single-shot, with a pulse duration of about 50 ps, much longer than all characteristic time scales, and much longer than for the measurements displayed in Fig.~\ref{fig:Fig_3}. Fig.~\ref{fig:Fig_4}~(a,b) shows a typical k-space emission measured on sample 2 in the RDSOC and parabolic regimes, respectively (log colormap). One can observe the presence of either one maxima or two maxima of emission in k-space. The k-space distance between the two peaks of the condensate emission $\Delta k=2k_0$ versus voltage is shown in Fig.~\ref{fig:Fig_4}~(c) (black dots) and is in full agreement with the theoretical expectations (blue line, See Section~2 in SM). Fig.~\ref{fig:Fig_4}~(d,e) shows the real space emission for two different shots, in the RDSOC regime, above threshold. We plot the emission of one linear polarization, for which the contrast is expected to be 100\% (see Methods, "Analytical description..."), and stripes are now extremely clear. Stripes measured from the full density on sample "3", with a lower contrast, are shown in SM Section~3. \textcolor{black}{Stripes are the result of the interference between the two k-components of the condensate. Spatial phase fluctuations of these components would destroy this interference, and would result in the presence of a large number of dislocations. Our stripe observation therefore directly proves that the condensate is spatially coherent at least over the region where the stripes are observed, without the need of any extra measurements}.
Now comparing Fig.~\ref{fig:Fig_4}~(d) end (e), one can see that the two realizations are different, with the most intense stripes being approximately in anti-phase with each other (the spatially-averaged values of the phase are -0.47$\pi$ and +0.48$\pi$, respectively). We access  the phase ($\Phi_1-\Phi_2$), characterizing the position of the stripes, and their contrast by making a Fourier transform of the spatial images. Fig.~\ref{fig:Fig_4}~(f) shows histograms of phase distributions over 500 realizations corresponding to a high-density condensate regime for the two left panels, and to a condensate density approximately 4 times smaller for the right panel. The phase in the high density regime is uniformly distributed in both sample "2" and sample "3", which suggests the absence of pinning and fully spontaneous breaking of translational invariance. On the contrary, the low-density regime shows a preferential phase corresponding to a partial pinning of stripes in different realizations. One should note that all the data are taken close to threshold. They correspond to a change in the pump density by only 10-20\%.  Another view of the same effect is plotting the contrast of the integrated image versus the number of integrated shots $N$ (up to 500), see Fig.~\ref{fig:Fig_4}~(g). In the high-density regime, the contrasts measured on both sample "2" and "3" decay approximately as $1/\sqrt{N}$, as expected for completely random phases (see Section~4 in SM). This spontaneous translational symmetry breaking appears to be insensitive to the random disorder present in the samples, which is exactly what one expects for superfluid stripes. 

A superfluid flowing in a disorder potential has its density affected by this disorder. On the contrary, its phase remains globally unaffected, because the phase of a superfluid is directly related to its velocity $\Vec{v}=\hbar\nabla\arg\psi/m$, and the absence of backscattering means that the velocity field, and therefore the phase, remains unperturbed thanks to superfluidity. The freedom of the phase of the condensate with stripes in presence of the disorder potential is therefore equivalent to a frictionless propagation of a superflow.

On the contrary, for lower condensate densities the contrast does not decrease further after averaging of about 10 realizations, which means that stripes choose a preferential location. As one can qualitatively expect, a superfluid behavior of the stripes is favored by a large density and large interaction energy comparable with the disorder amplitude and with the kinetic energy involved in the stripe phase, but also in the dynamical process of condensate formation. This partial pinning just above threshold also explains why we can still observe stripes in multiple-shot experiments with averaging over 500 shots.

We simulate polariton condensation under non-resonant pumping using the hybrid Boltzmann-Gross-Pitaevskii equation with lifetime, energy relaxation, saturated gain \cite{Wertz2012}, spin-anisotropic polariton-polariton interactions, and disorder. All details on the model and parameters used are given in  Methods. Fig.~\ref{fig:Fig_4}~(h,i) shows the phase histogram and contrast decay versus $N$ for two different condensate densities. They reproduce both the pinning effect of the disorder at low densities and the random positioning of the stripes when the interaction energy in the condensate becomes comparable with the disorder magnitude. A discussion about the expected magnitude of interactions in this type of structure is given in the Section~5 in SM.
In Section~6 in SM, we present simulations performed for \textit{continuous-wave} pumping, where we compute the dispersion of elementary excitations (bogolons) of the stripe phases, as previously done for equilibrium atomic condensates \cite{li2013superstripes} \textcolor{black}{(we do not measure this dispersion experimentally, because the signal from the condensate is too strong)}. The difference is that here we model a system with pump, decay and possibly disorder. These simulations confirm the existence of gapless linearly dispersing bogolons that are characteristic for a superfluid. We point out that such behaviour cannot be achieved for striped condensates composed of negative mass particles \cite{trypogeorgos2025emerging}, which form bright solitons.

\section*{Observation of quantum vortices in a supersolid}

\begin{figure}
\centering
\includegraphics[width=0.8\linewidth]{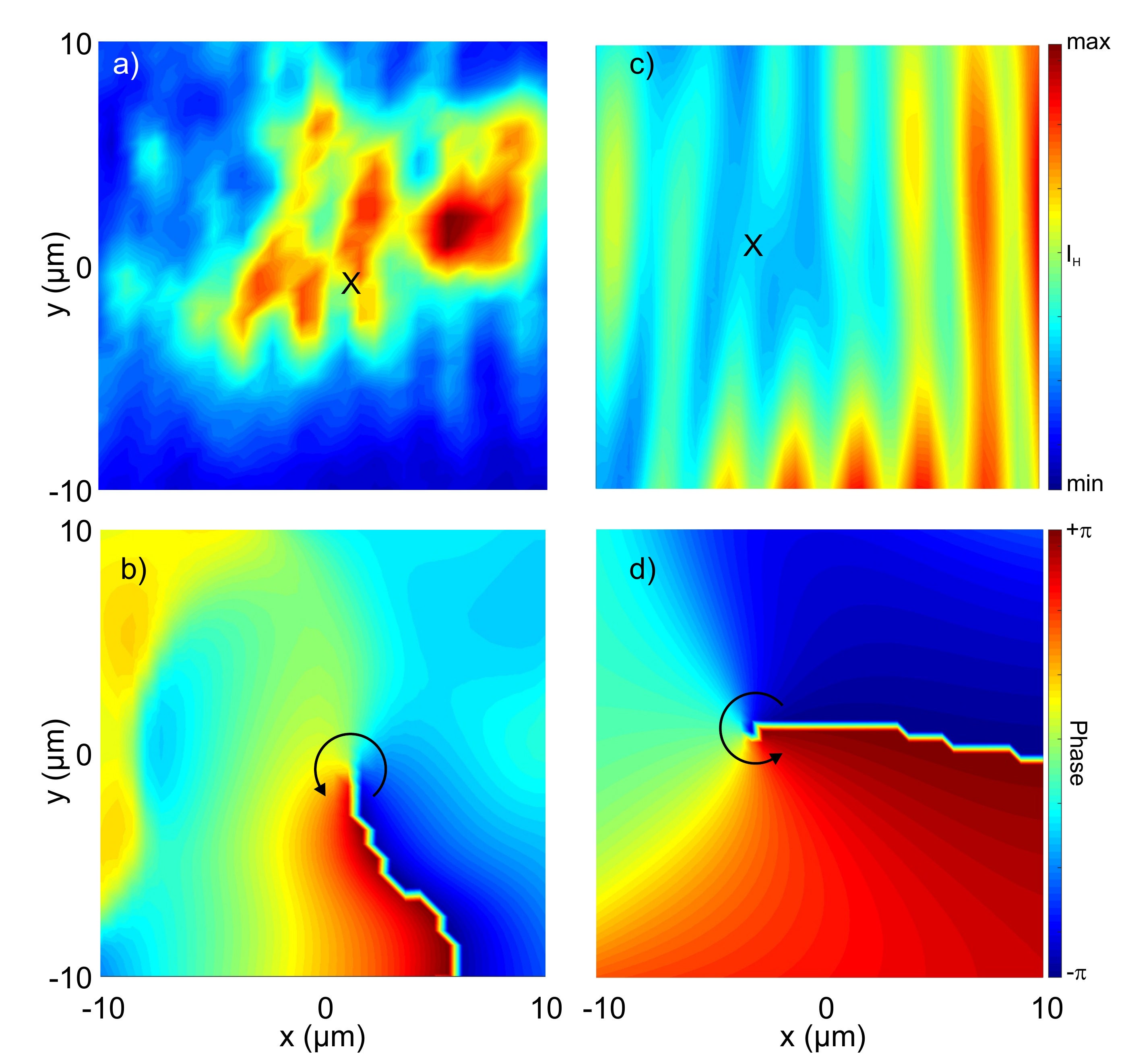}
\caption{\small \label{fig:Fig_5} \textbf{Dislocations and relative phase vortices in single-shot experiments and simulations.} a,c) A single-shot image with the stripes showing a dislocation (marked by cross). b,d) The extracted phase of the fringes exhibits a vortex singularity. Arrow marks the phase gradient. (a,b) - Experiments, (c,d) - simulations.  }
\end{figure}

Another effect which affects the stripe patterns is the possibility of formation of different regions with different relative phases, leading to the formation of vortices. Random creation of vortices takes place during the condensation process via the Kibble-Zurek mechanism \cite{weiler2008spontaneous,solnyshkov2016kibble}. In an ultracold atomic supersolid, they were recently generated by putting the condensate in rotation~\cite{casotti2024observation}. In Fig.~\ref{fig:Fig_5}~(a,b) we show an example of such a topological defect, which spontaneously appears in the condensation process as a dislocation in the pattern of fringes observed for a particular single-shot image (Fig.~\ref{fig:Fig_5}~(a)), and the associated phase extracted from the Fourier transform (Fig.~\ref{fig:Fig_5}~(b)), exhibiting a characteristic vortex close to the origin of the coordinates (the uncertainty on the vortex position is of the order of the period of the stripes).
The results of simulations showing a similar effect are shown in Fig.~\ref{fig:Fig_5}~(c,d). Section 7 of SM presents movies showing the time evolution of stripes during the condensation process. This allows one to observe quantum vortex (dislocation) creation and motion, as well as the dynamic behavior of stripes, even in the presence of disorder. In experiments, vortices are observed at random positions, so they are not directly related to or pinned by the disorder potential. They are observed in about 10\% of experiments, with the probability depending on the pumping power. They also affect the decay of contrast with $N$ (See Section~4 in SM). The exact scaling of their number (density) versus the pumping power which is the quench parameter will be the topic of a future work.

In summary, we demonstrated strong light-matter coupling in an optical microcavity with embedded LC and photostable MeLPPP polymer. We observed polariton condensation first in the regime where photonic bands of the cavity are well separated and we demonstrated the tunability of the condensate polarization by either the external voltage or the pump polarization. In the RDSOC regime, non-equilibrium condensate forms simultaneously in the negative mass state of the dispersion, being a signature of the repulsive interaction with the excitonic reservoir, and in two degenerate minima of the dispersion. The real space distributions of polarization and intensity reveal the presence of stripes. By single-shot experiments, we observe that polarization stripes are located randomly from one condensate realization to another, demonstrating the spontaneous character of the translational invariance symmetry breaking. The robustness of this process against the intrinsic disorder potential of the structure, together with the detection of quantum vortices demonstrates the superfluid behaviour of the stripe phase, confirming the supersolid nature of the observed phase.
Our work opens new perspectives for quantum fluids of light physics in presence of SOC and beyond, in topologically non-trivial systems, both Hermitian and non-Hermitian, at room temperature.

\begin{methods}
\renewcommand{\figurename}{Extended Data Figure}
\setcounter{figure}{0}   

    \textit{Sample preparation.} 
For all samples, the cavity consists of two distributed Bragg reflectors (DBRs) composed of 15 and 13 SiO${_2}$/Ta${_2}$O${_5}$ layers, with a central wavelength at 490~nm. The DBRs were deposited on 30~nm ITO electrode layers on quartz substrates with a flatness of $\lambda$/20(@633~nm). Both DBRs were spin-coated with 35~nm MeLPPP layers covered with protective Al$_{2}$O$_3$ layers of 20~nm thickness. Antiparallel orienting layers (SE-130, Nissan Chem., Japan) were deposited on both substrates using the spin-coating method. The wedge cavity was achieved using glass spacers with sizes of 1.5~--~2~$\mu$m placed between the substrates. The cavity was filled with a liquid crystal mixture, LC2091$^{\ast}$ (refractive indices $n_o\approx1.57$ and $n_e\approx1.98$, for the sodium line 589~nm), in the nematic phase by capillary action.

\textit{\label{sec:sm4}Experimental details.}
Extended Data Figure~\ref{fig:exp_setup} presents the optical experimental setup used in multiple-shot experiments. The optical measurements were performed at room temperature. Reflectivity measurements employed a microscope objective with 50x magnification, numerical aperture NA = 0.65 for both the excitation and the collection of light. The spot had a diameter of approximately 10~$\mu$m on the sample. In photoluminescence measurements, an additional lens \textit{f}~=~100~mm was used to focus the excitation laser onto the 20~$\mu$m spot on the surface of the sample. The 181~fs pulsed laser of 425~nm wavelength and repetition rate of 500~Hz was used for non-resonant excitation. Depending on the selected lenses (k-space or r-space) and the light source, the setup allowed measurements in both reciprocal and real space at the same position on the sample. Two sets of a quarter-wave plate, a half-wave plate, and a linear polarizer were used to control the polarization of the excitation and collected signal. To tune the dispersion relation of the cavity modes, the LC microcavity was addressed by an AC waveform generator with a 1010~Hz square signal and varying amplitude of 0~--~20~V. Paths of white light are indicated by yellow lines, the laser light path is marked with a blue line, and the signal from the sample is represented by a cyan line.

Single-shot images were collected using a modified experimental setup compared to the Extended Data Figure~\ref{fig:exp_setup}. The LC microcavity was addressed by an AC waveform generator with 1000 Hz square signal that is phase-synchronized with the excitation laser pulses. The condensate was non-resonantly excited with pulses at 400~nm wavelength.  Coupling the 150~fs pulses into a 25~$\mu$m multi-mode fiber allows us to temporarily stretch them to several tens of ps (about 50~ps) and homogenize the beam profile and the polarization. The excitation light was focused onto the sample with a 20x objective (NA~=~0.42). The condensate emission was collected with a 10x objective (NA~=~0.30) and detected in a real-space configuration by a camera with a linear polarizer placed in front. 

\begin{figure*}[tbh]
\centering
\includegraphics[width=1.0\linewidth]{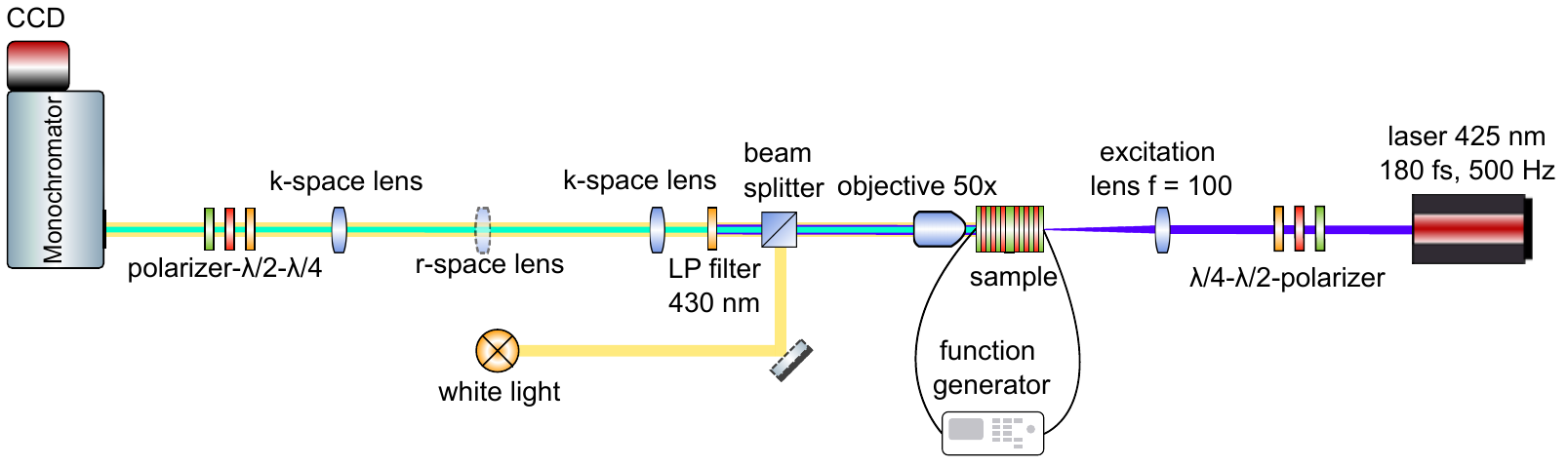}
\caption{\label{fig:exp_setup} \textbf{Experimental setup scheme.}}
\end{figure*}

\textit{\label{sec:sm1}MeLPPP absorption and complex refractive index.}
Large inhomogeneous broadening of the excitonic peak plays a significant role in the theoretical treatment of the results. In order to estimate this broadening, we measured the absorption spectrum of a MeLPPP layer on the DBR and made a fit of the main excitonic peak by Gaussian function $\exp{[-(E-E_X)^2/2 \sigma^2]}$, with $E_X=2.715$~eV center of the peak and $\sigma=35$~meV standard deviation of exciton energy distribution (Ext. Data Fig.~\ref{fig:sm_1}(a)). Since we will be considering only strongly negatively detuned photonic modes, we do not need to take into account higher energy absorption peaks.

\begin{figure}[tbh]
\centering
\includegraphics[width=0.95\linewidth]{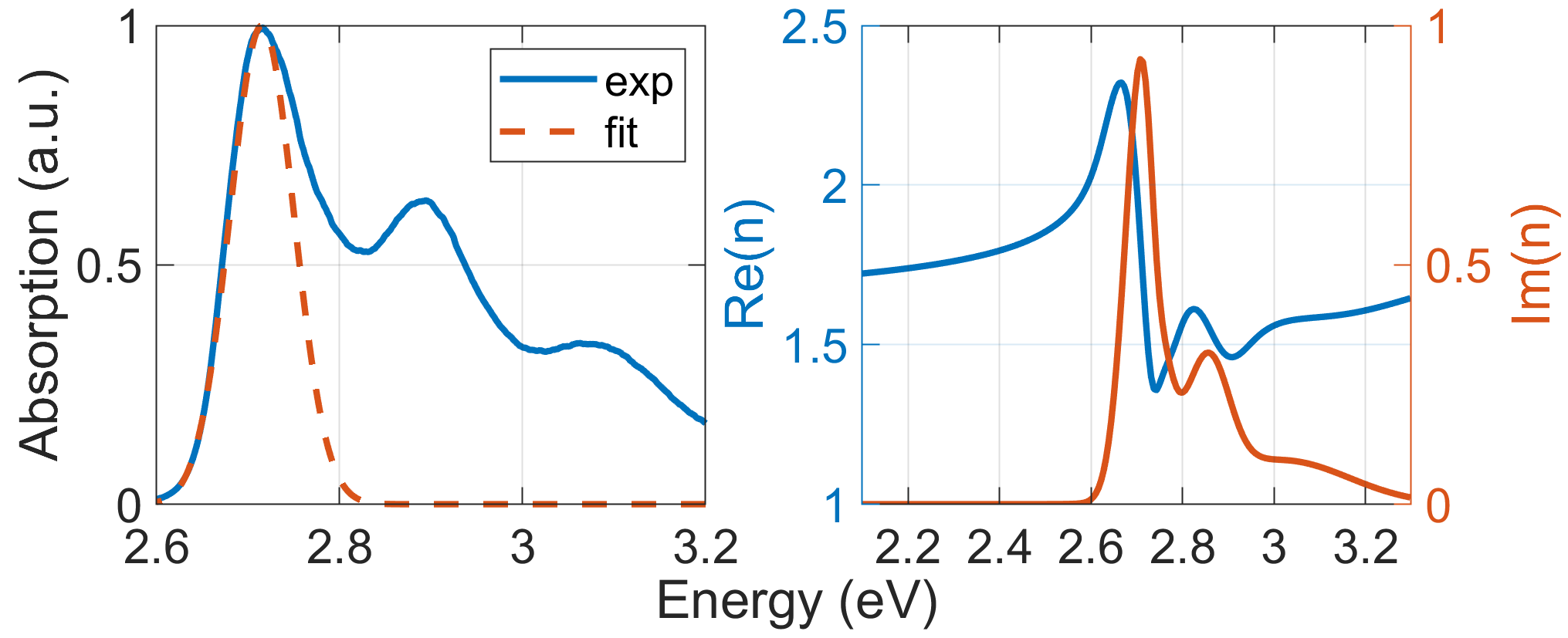}
\caption{\label{fig:sm_1} \textbf{MeLPPP absorption and complex refractive index.} Left: Experimentally measured absorption of a single MeLPPP layer on the DBR (solid blue line); the fit of the main absorption peak by Gaussian function representing inhomogeneous broadening of excitons (dashed red line); Right: real (Re($n$), blue line) and imaginary (Im($n$), red line) parts of complex refractive index obtained by variable angle spectroscopic ellipsometry.}
\end{figure}

Also, we show here the MeLPPP complex refractive index real and imaginary parts (Ext. Data Fig.~\ref{fig:sm_1}(b)) which will be used in the next section for the Berreman method simulation of microcavity structure. The background refractive index of MeLPPP extracted from Re($n$) dependence is $n_{BG}=1.7$.

\textit{\label{sec:sm2}Photonic mass.}
Since the mode 15H is tuned over a large range of energies in Fig.~2 (LPB detuning between -188 meV and -84 meV), it is crucial to estimate how the photonic mass of the mode changes in the tuning process. In order to do this, we first reproduce the band structure of the microcavity with the use of the Berreman method~\cite{berreman1972optics}. We simulate the structure sketched in Fig.~1(b) and tune the length of the LC layer and the angle of LC molecules in order to obtain perfect correspondence between measurement and simulation shown in Ext. Data Fig.~\ref{fig:sm_2}. In the experiment, the PL was measured in H polarization, which is why the H mode is well-visible for any voltage applied to the microcavity, while the V mode appears only when it couples to the H mode strongly enough (Ext. Data Fig.~\ref{fig:sm_2}(a)). The numerical simulation (Ext. Data Fig.~\ref{fig:sm_2}(b,d)) is represented by the first Stokes parameter of the system (linear polarization degree) in order to easily distinguish between H and V modes, appearing as blue and red lines, respectively. After achieving correspondence between the experiment and numerical simulation, we calculate once again the dispersion for the same structure, except that now the MeLPPP refractive index is constant and equal to background one ($n=n_{BG}$), thus eliminating the exciton from the calculation and making the system purely photonic. Then, we extract the mass and the position of the photonic mode and plot them in Ext. Data Fig.~\ref{fig:sm_2}(e). The range of photonic mode detunings $E_{ph}-E_{ex}$ corresponds to the LPB detunings from Fig.~2(d). One can see that the total change of the photonic mass is approximately $19\%$. This simulation proves that the much bigger mode effective mass change observed in Fig.~2 can be explained solely by exciton-photon coupling. 

\begin{figure}
\centering
\includegraphics[width=\linewidth]{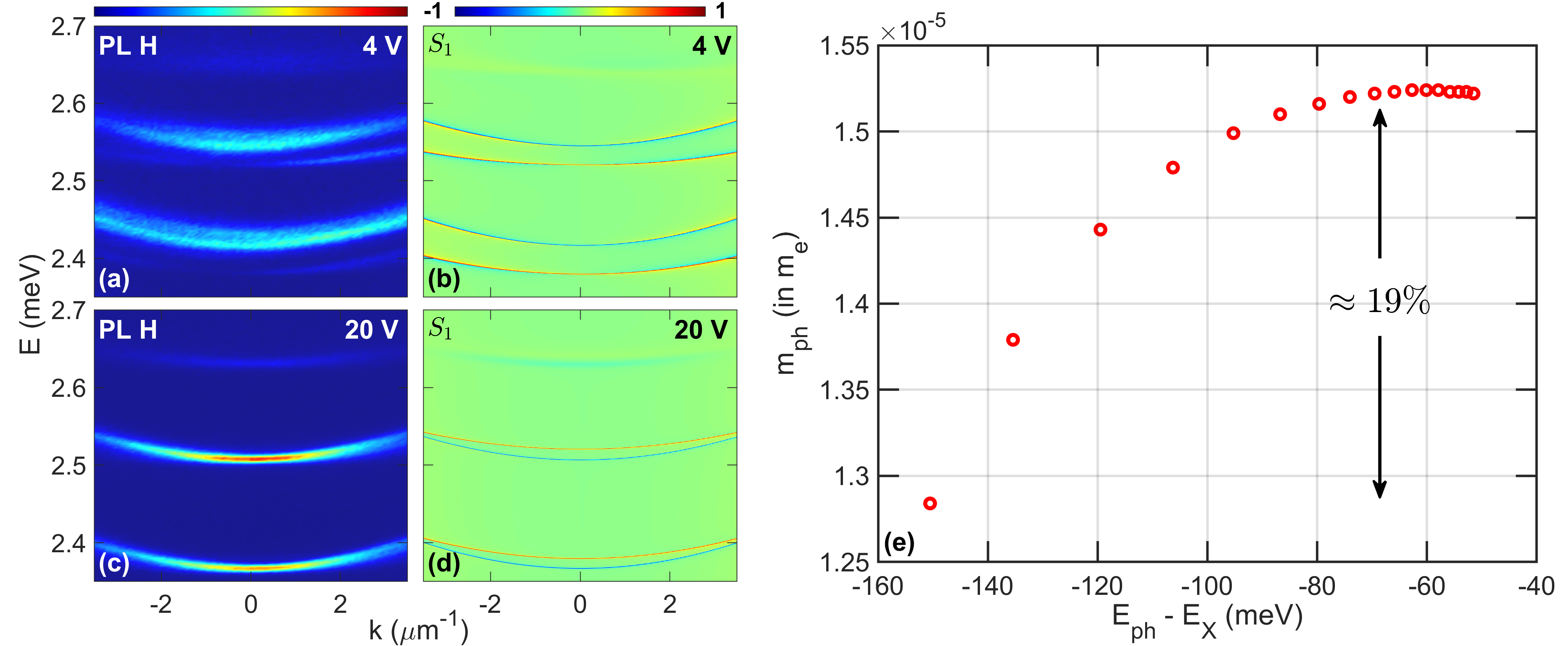}
\caption{\label{fig:sm_2} \textbf{Photonic mass extraction.} (a,c) Experimentally measured PL in H polarization and (b,d) numerically calculated first Stokes parameter (degree of linear polarization) with the use of the Berreman method for the voltage (a,b) 4 V and (c,d) 20 V applied to the microcavity. (e) Extracted from simulation, the photonic mass as a function of photonic mode detuning. The photonic detunings range corresponds to the LPB detunings range from Fig.~2(d). The total photonic mass change is $19\%$, proving that the mass change of $76\%$ shown in Fig.~2(d) can be explained only by exciton-photon coupling.}
\end{figure}

\textit{\label{sec:sm3}The Tavis-Cummings model.}
The Tavis-Cummings model describes the system consisting of a single cavity mode coupled to $N$ two-level systems, excitons in our case. Usually, the model utilizes identical excitons, while in our case the inhomogeneous broadening of the excitonic peak plays an important role in quantitative correspondence with experiment. The model reads:
\begin{eqnarray}
    \label{TC_Ham}
    H &=& \hbar (\omega_{ph} - i \gamma_{ph}) a^\dagger a + \sum_{j=1}^N \hbar (\omega_{ex,j} - i \gamma_{ex}) \sigma_j^+ \sigma_j^-\nonumber \\
    &+&\hbar g \left( a \sigma_j^+ + a^\dagger \sigma_j^- \right),
\end{eqnarray} 
with $\hbar \omega_{ph} = \hbar \omega_{ph,0} + \hbar^2 k_x^2/2 m_{ph}$ energy of a cavity defined by $\hbar \omega_{ph,0}$ energy at $k_x=0$ and $m_{ph}$ cavity photon mass, $\gamma_{ph}$ cavity mode decay, $\hbar \omega_{ex,j}$ energy of $j$-th exciton taken from inhomogeneously broadened Gaussian distribution, $\gamma_{ex}$ non-radiative decay of a single exciton, $\hbar g$ exciton-photon coupling strength, $a^\dagger$ and $a$ cavity mode creation and annihilation operators, respectively, $\sigma_j^+ = \ket{e_j} \bra{g_j}$ and $\sigma_j^- = \ket{g_j} \bra{e_j}$ $j$-th two-level system raising and lowering operators, respectively. The total Rabi splitting can be estimated as $\Omega=\hbar g \sqrt{N}$ for large exciton-photon detunings.

For every voltage, we extracted the photonic mode energy $\hbar \omega_{ph}$. Since the characteristic pump spot size ($w_{pump} \approx 20 \ \mu$m) is much larger than the estimated transverse cavity photon coherence length ($l_{coh} \approx 1.5 \ \mu$m), we consider $N_{ens} = w_{pump}^2/l_{coh}^2$ different realizations of Hamiltonian~\eqref{TC_Ham}: each realization differs by excitonic disorder (ensemble of $\hbar \omega_{ex,j}$) and coupled photon. Using extracted photonic data, we diagonalize each realization of the Hamiltonian~\eqref{TC_Ham} and for each realization we extract the eigenmode with the highest photonic contribution. Next, we sum up the PL signal originating from these eigenmodes for all wavevectors. After applying this procedure, we obtain a total PL signal similar to the experimental one. We then extract the mass and energy of the numerically obtained LPB, exactly like we do with the experimentally measured PL intensity maps, and plot the final results in Fig.~2(d). As one can see both LPB detuning and mass obtained by numerical simulation correspond well to the experimental data when we choose Rabi splitting of $\Omega=93$~meV. The number of excitons used for every realization of Hamiltonian~\eqref{TC_Ham} is $N_X=200$. In reality, the number of excitons per each coherent spot $l_{coh}^2$ should be much bigger, but increasing this number very quickly makes the Hamiltonian diagonalization computationally heavy. That is why we made sure that the simulation results slowly converge when we increase $N_X$ up to 2000, then we performed all the simulations for $N_X=200$ and included the difference as a method error shown by red error bars in Fig.~2(d). The decay values used in simulations are $\hbar \gamma_{ph}=3$~meV and $\hbar \gamma_{ex}=0.1$~meV~\cite{plumhof2014room}.

\textit{Anticrossing.}
\begin{figure}
    \centering
    \includegraphics[width=1.0\linewidth]{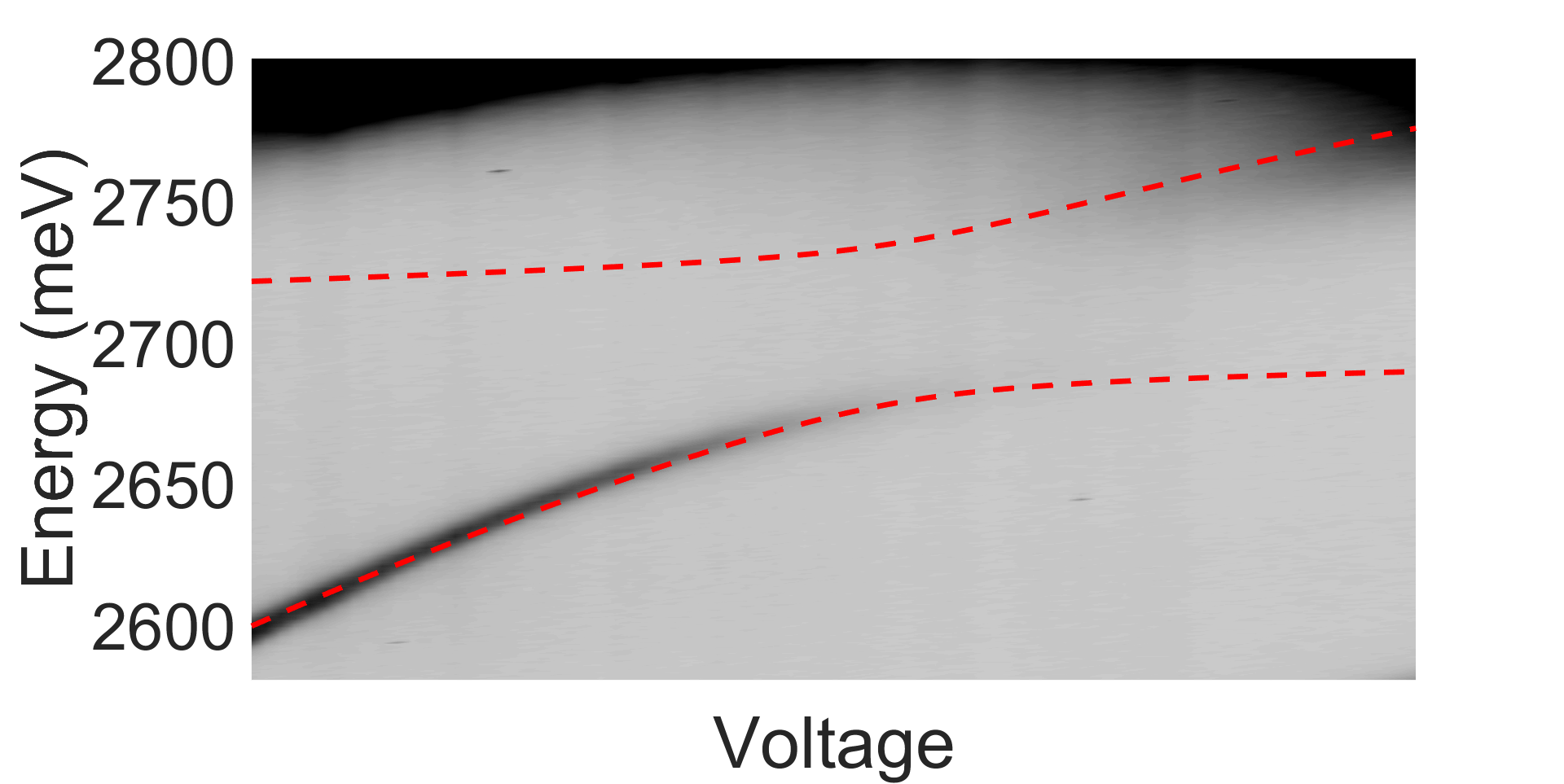}
    \caption{\textbf{Anticrossing of exciton-polariton modes.} White light transmission of the sample "3" versus the applied voltage. The dashed lines are computed from a two damped oscillator model as described in the text.}
    \label{Anti}
\end{figure}
Extended Data Figure ~\ref{Anti} shows the white light transmission spectrum versus voltage measured on a sample "3", similar to the two samples studied in the main text. However, the H-mode here displays a clear anti-crossing with the exciton resonance. It was shown by Kavokin and Kaliteevski \cite{kavokin1995excitionic} in 1995, from Maxwell equations that the optical response of a microcavity can be well described by a two damped oscillator model. Such model was used by Savona et al. \cite{savona1995quantum} to assess the weak to strong coupling transition. Within this model, the energy splitting between modes reads $\sqrt{(E_{ph}-E_x+i(\hbar\gamma_{ph}-\hbar\gamma_{ex}))^2+(\hbar g)^2}$. The fit uses $\hbar g$ = 90~meV, in agreement with the values found using the more accurate Tavis-Cummings model and $\hbar\gamma_{ex}-\hbar\gamma_{ph}=$~70~meV.
At zero exciton-photon detuning, the same expression reads $\hbar\sqrt{g^2-(\gamma_{ph}-\gamma_{ex})^2}$. 
If the broadening is larger than 90 meV, then it leads to the weak coupling regime at zero detuning, where the real parts of mode energies are degenerate. This zero detuning condition is what is nowadays called an exceptional point. When the detuning is non-zero, the square root above acquires a non-zero real part and the eigenmode energy splitting 
initially scales as a square root of the detuning. It means that everywhere, but at zero detuning eigenstates are mixed exciton-photon states. However, Savona proposed that in order to be significant, the strong coupling  should require that the real part splitting exceeds the sum of the imaginary parts corresponding to the mode linewidths. It means that two resonant peaks should be distinguishable in the optical spectroscopy response of the cavity.

This condition is realized by far when the exciton-photon detuning $|E_{ex}-E_{ph}|$ exceeds the difference between the mode linewidths $|\hbar(\gamma_{ph}-\gamma_{ex})|$.
The system can therefore be weakly coupled at zero detuning, and strongly coupled at non-zero detuning, which leads to the very strong mass change experimentally observed and well described by the Tavis-Cummings model which is the most appropriate to describe systems with large inhomogeneous broadening. 

The Rabi splitting value controls the level of mixing which can be realized at the detuning where strong coupling takes place. At a detuning of -200~meV, as the one for which the condensation is taking place, the exciton fraction is between 5 and 10\% and only marginally depends on whether the system is weakly or strongly coupled at zero detuning, (i.e. if the broadening is 70 or 95~meV, for instance).

\textit{Interference measurements.}
Extended Data Figure~\ref{fig:coher} shows the interferogram of a polariton condensate measured at excitation intensity of around $1.2P_{th}$ with Michelson interferometer. The experiment was not performed simultaneously with the one presented in the main text and was using a slightly different excitation scheme. We non-resonantly excite the condensate using  a 6~ns pulsed laser at a 425~nm wavelength in a single-shot regime. Interferometry data were acquired at 0~V, so in the regime of parabolic dispersion (RDSOC off), and with a pump spot of 20~$\mu$m. In Figure~\ref{fig:coher} the interference fringes are observed over the whole area of the condensate in the real space, demonstrating a build-up of the long-range spatial coherence, which is a hallmark of the exciton-polariton condensation.

\begin{figure}
\centering
\includegraphics[width=0.6\linewidth]{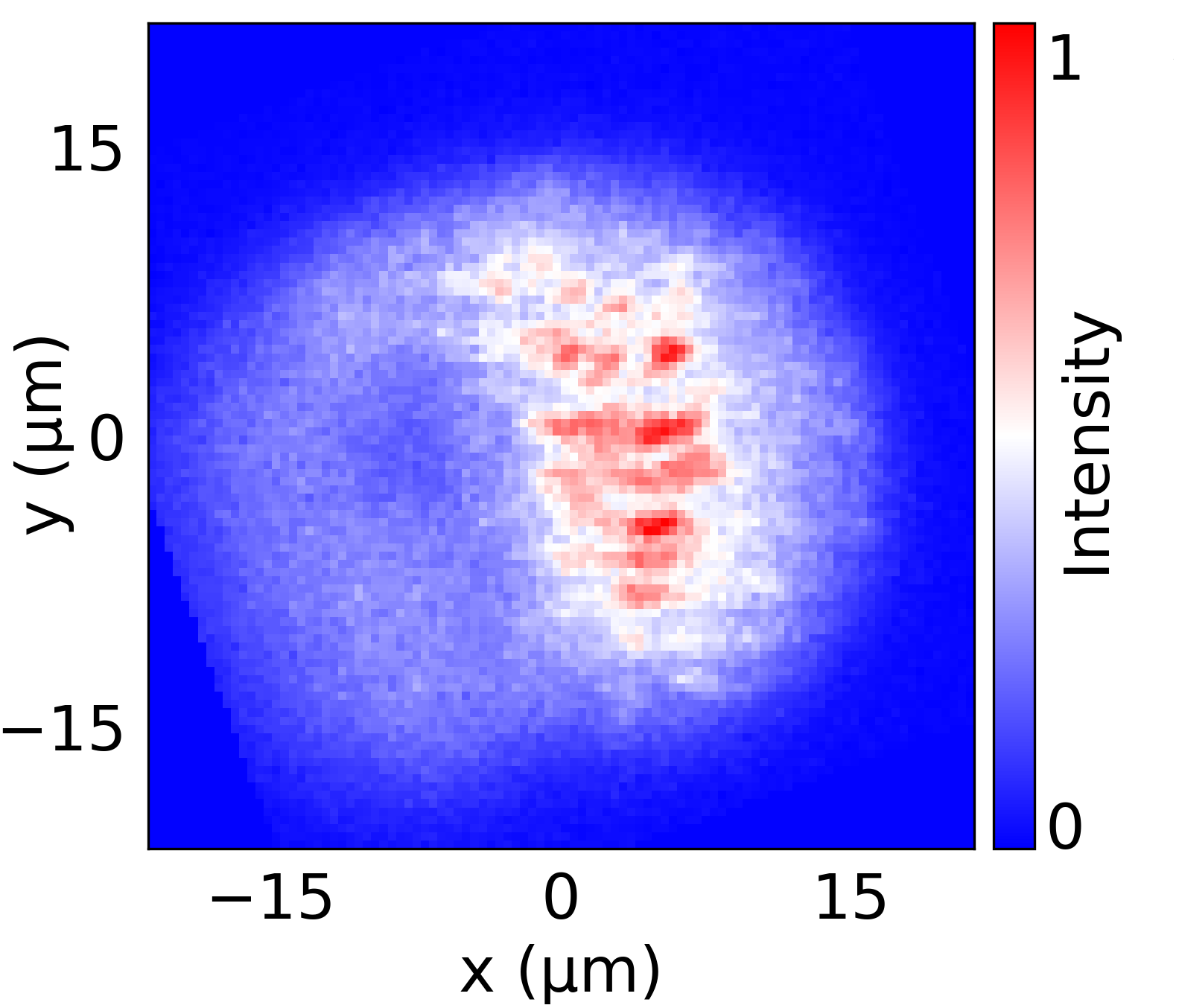}
\caption{\label{fig:coher} \textbf{Single-shot interferogram of a polariton condensate at zero delay for sample 1.} Similar results were also obtained for sample 2.}
\end{figure}

\textit{Analytical description of the stripes and their contrast.}
The wave function describing a spinor wavefunction with two contributions corresponding to the eigenstates at two opposite wave vectors $\pm k_0$ can be written in circular basis as:
\begin{equation}
\left| \psi  \right\rangle  =\frac{1}{2} \left( {\begin{array}{*{20}{c}}
{\cos \frac{\theta }{2}}\\[2pt] 
{\sin \frac{\theta }{2}}
\end{array}} \right){e^{ - ik_0x-i\phi_1}} + \frac{1}{2}\left( {\begin{array}{*{20}{c}}
{\cos \frac{{\pi  - \theta }}{2}}\\[2pt] 
{\sin \frac{{\pi  - \theta }}{2}}
\end{array}} \right){e^{ + ikx_0-i\phi_2}}
\end{equation}
where $\theta$ is the polar angle of the Stokes vector, whereas the azimuthal angle is fixed by the polarization of the coupled modes (H and V). The phases $\phi_1$ and $\phi_2$ of the two orthogonal components are chosen randomly due to the spontaneous symmetry breaking. Rewriting the expression allows one to make appear explicitly a random phase responsible for the position of the stripes, and another one (an overall global phase), which can be detected in coherence measurements via interference (as in Extended Data Figure~\ref{fig:coher}, for example):
\begin{equation}
\left| \psi  \right\rangle  =\frac{1}{2}e^{i\zeta}\left( \left( {\begin{array}{*{20}{c}}
{\cos \frac{\theta }{2}}\\[2pt] 
{\sin \frac{\theta }{2}}
\end{array}} \right){e^{ - i(k_0x-\eta)}} + \left( {\begin{array}{*{20}{c}}
{\cos \frac{{\pi  - \theta }}{2}}\\[2pt] 
{\sin \frac{{\pi  - \theta }}{2}}
\end{array}} \right){e^{ + i(kx_0-\eta)}}\right)
\end{equation}
where the global phase is $\zeta=-(\phi_1+\phi_2)/2$, and the relative phase controlling the position of the stripes is $\eta=-(\phi_1-\phi_2)/2$.

The corresponding density reads
\begin{equation}
\left\langle {\psi }
 \mathrel{\left | {\vphantom {\psi  \psi }}
 \right. \kern-\nulldelimiterspace}
 {\psi } \right\rangle  = \frac{1}{2} + \frac{1}{2}\cos 2(k_0 x-\eta)\sin \theta 
 \end{equation}
It exhibits oscillations, whose contrast is $U=\sin\theta$ and whose position is controlled by the randomly chosen phase $\eta$. This expression gives a direct relation between the circular (or linear) polarization degree with the contrast of the stripes observed in total intensity. It also allows to understand the decrease of the contrast in multiple-shot integrated experiments, where the position of stripes in each shot is random (see Supplementary Materials).

Optical experiments allow to measure the intensity of a single linear polarization, with its amplitude obtained from the circular components as $\ket{\psi_H}=(\psi_1+\psi_2)/\sqrt{2}$. This allows to strongly improve the contrast, as required for single shot measurements:
\begin{eqnarray}
    I_H&=&\braket{\psi_H|\psi_H}=\sin^2\left(\frac{\pi}{4}+\frac{\theta}{2}\right)\cos^2\left(k_0 x -\eta\right)\nonumber\\
    &=&\sin^2\left(\frac{\pi}{4}+\frac{\theta}{2}\right)\left(\frac{1}{2} + \frac{1}{2}\cos 2(k_0 x-\eta)\right)
\end{eqnarray}
The contrast of this signal is always 100\% (at least without the experimental background noise), no matter the value of $\theta$, contrary to the stripes observed directly in $\braket{\psi|\psi}$, while the origin and the position of the stripes are exactly the same, governed by the phase $\eta$, as can be seen from the cosine function.

\textit{Numerical simulations.}
To describe the polariton condensation in our system, we solve the following equation \cite{Shelykh2006,Solnyshkov2014pra,Wertz2012}:
\begin{eqnarray}
i\hbar \frac{{\partial {\psi _ \pm }}}{{\partial t}} &=& \left( {1 - i\Lambda } \right){\hat T_ \pm }{\psi _ \pm }  + g_1 |\psi_{\pm}|^2\psi_{\pm}\nonumber\\
&+&g_2 |\psi_{\mp}|^2\psi_{\pm}+i\gamma(|\psi|^2){\psi _ \pm } + U{\psi _ \pm }\nonumber\\
&+&\Delta\psi_\mp+\chi.
\label{gpe}
\end{eqnarray}
Here, $\hat{T}_\pm$ is the kinetic energy operator containing terms accounting for several effects: the dispersion relation of polaritons with mass $m$, the gauge potential contribution of the RDSOC $-2i\alpha \cdot \partial\psi_\pm/\partial x$, and the energy relaxation mechanisms~\cite{Pitaevskii58} via $\Lambda$, with zero energy level set at the bottom of the dispersion. $U=U_D+U_R$ is the potential, where $U_D$ is the disorder potential (correlation length $7~\mu$m, rms fluctuation 0.3~meV) and $U_R=U_0\exp(-r^2/2\sigma^2)$ is the repulsive potential of the reservoir. $\gamma(|\psi|^2)$ is the term combining decay and saturated gain, $\chi$ is the noise describing the spontaneous scattering from the excitonic reservoir. The saturated gain is given by $\gamma(\left|\psi\right|^2)=\gamma_0(n_R)\exp(-n_{tot}/n_s)$
with $n_{tot}=\int|\psi|^2\,dxdy$ the total particle density, $n_s$ the saturation density, and $n_R$ the exciton reservoir density. Both the gain and the reservoir potential account for the duration of the pulsed excitation. $g_1$ and $g_2$ are the polariton-polariton interaction constants (same-spin and opposite-spin, respectively, with $g_2<0$ and $|g_2|\ll g_1$ for negative detunings~\cite{Vladimirova2010}). We take $g_2=0$, and $g_1$ such as to have a maximal value of blue shift of 1~meV above threshold. We solve Eq.~\eqref{gpe} numerically using the parameters of the polariton dispersion determined from the experimental results presented in Fig.~3 of the main text. Other parameters were taken as $\Lambda=0.02$, $U_0=0.1$~meV, $\sigma=18~\mu$m. 
These parameters were used in simulations shown in Fig.~4 of the main text.

A specificity of this platform with respect to atomic condensates is the mass difference $\gamma$ between two linearly polarized modes coupled by the RDSOC. It ensures the presence of stripes even at zero detuning $\Delta$ with a contrast $M/2m$ (see Eq.~(3) from main text). However, stripes disappear at a particular non-zero value of detuning $\Delta = -\gamma k_0^2$.

\end{methods}

\bibliographystyle{naturemag}
\bibliography{apssamp}

\begin{addendum}
\item This work was supported by the National Science Centre grants 2019/35/B/ST3/04147, 2019/33/
B/ST5/02658 and 2022/47/B/ST3/02411, 2023/51/B/ST3/03025, and the Ministry of National Defense Republic of Poland Program -- Research Grant MUT Project 13-995 and MUT University grant (UGB) for the Laboratory of Crystals Physics and Technology for the year 2021 and the European Union’s Horizon 2020 program, through a FET Open research and innovation action under the grant agreements No. 964770 (TopoLight) and EU H2020 MSCA-ITN project under grant agreement No. 956071 (AppQInfo). Additional support was provided by the ANR Labex GaNext (ANR-11-LABX-0014), the ANR program "Investissements d'Avenir" through the IDEX-ISITE initiative 16-IDEX-0001 (CAP 20-25), the ANR project MoirePlusPlus (ANR-23-CE09-0033), and the ANR project "NEWAVE" (ANR-21-CE24-0019). We thank Etsuki Kobiyama, Daniele Caimi and the team of the IBM Binnig and Rohrer Nanotechnology Center for support with the sample fabrication.
\item[Author contributions]
J.S., G.M., D.D.S., T.S., R.F.M., W.P., B.P. acquired funding; M.M.,, P.T., P.O., P.Ka. performed the experiments under the guidance
of B.P. and J.S.; P.Ku. synthesized liquid
crystal; R.Ma., P.M., and W.P. constructed and fabricated the LC microcavity;D.U., R.M., I.G. fabricated the DBR mirrors and the polymer layer with encapsulation and performed the basic optical characterization; M.F. and U.S. provided the polymer; D.D. performed single-shot coherence measurements;
P.Ko., D.D.S., and  G.M. performed theoretical analysis. All authors participated in the interpretation of
experimental data; J.S. and G.M. supervised the project;
P.Ko., M.M., G.M., D.D.S.,J.S. wrote the manuscript with input from all other
authors. M.M., P.Ko., D.D.S. and G.M. proposed the visualization of experimental
and theoretical data.
\item[Competing Interests] The authors declare that they have no
competing financial interests.
 \item[Correspondence] Correspondence
should be addressed to :  guillaume.malpuech@uca.fr (G.M.);\\ jacek.szczytko@fuw.edu.pl (J.S.).

\end{addendum}

\section*{Data availability statement}
The data generated in this study are available in the Open Science Framework (OSF) repository:
\verb|https://osf.io/hvpmn/?view_only=dea66a4b12b14ad0ac45451ec4205cc8|

\renewcommand{\thefigure}{S\arabic{figure}}
\renewcommand{\theequation}{S\arabic{equation}}
\renewcommand{\thesection}{\arabic{section}}

Supplemental Materials:\\Observation of a supersolid phase in a spin-orbit coupled exciton-polariton Bose-Einstein condensate at room temperature



\renewcommand{\figurename}{Figure}
\setcounter{figure}{0}

\section{Additional voltage for multiple-shot averaged experiment}

\begin{figure}[tbp]
    \centering
    \includegraphics[width=1.0\linewidth]{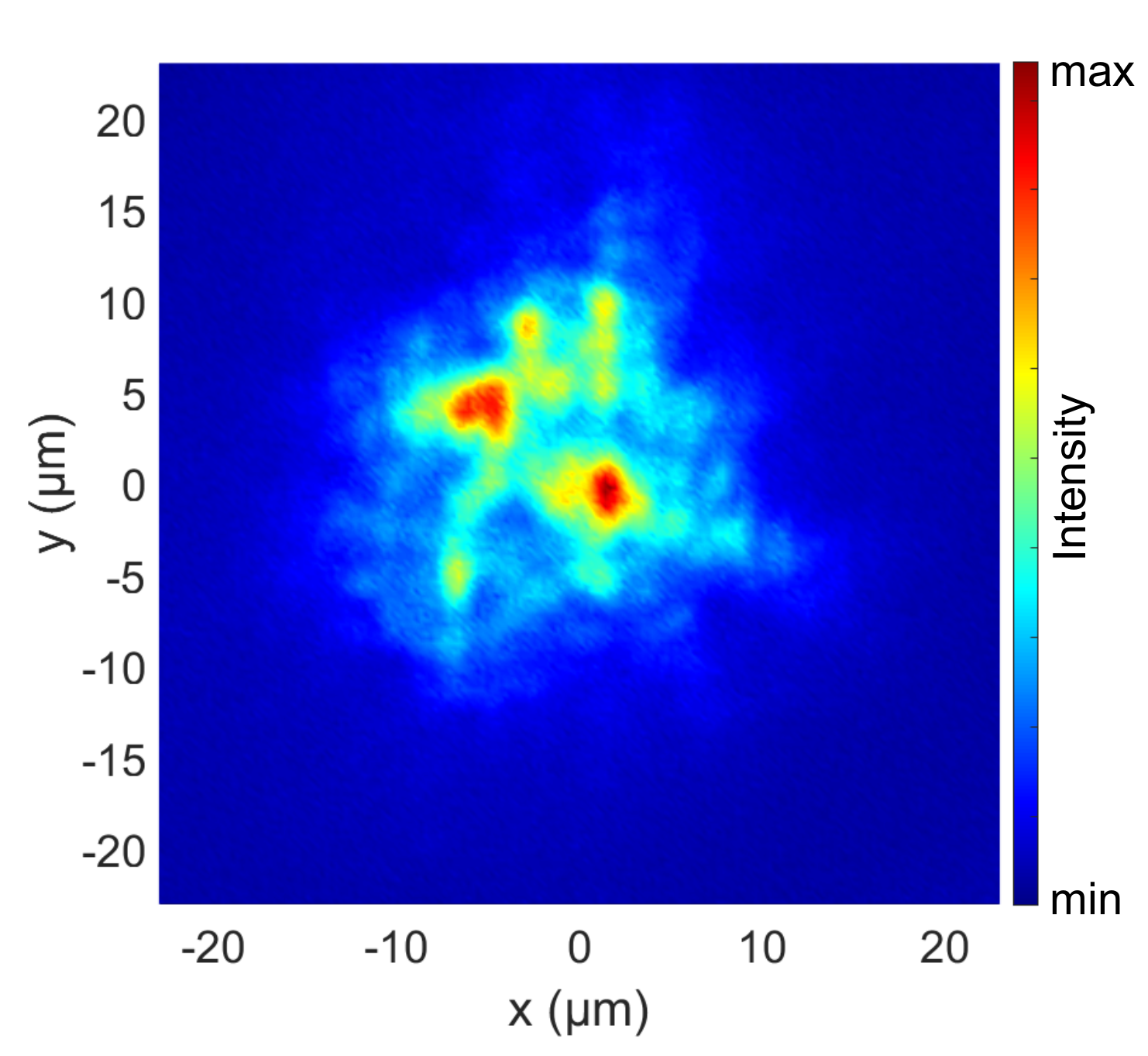}
    \caption{\textbf{Density stripes at a different voltage (sample "1").} Total intensity as a function of coordinates at 8.10~V. }
    \label{figSothervoltage}
\end{figure}

In the main text, we have presented a set of results for sample "1", corresponding to a limited set of voltages. In order to confirm that the stripes can also be observed at other voltages in this sample, here we present an additional figure. Figure~\ref{figSothervoltage} shows the stripes observed in the total density at a different voltage (8.10~V) in the same set of experiments as in Fig.~3 of the main text, with averaging over multiple pulses.

\section{Experimental study of the condensate in the reciprocal space}

The formation of a supersolid is a consequence of the condensation in a dispersion with two minima at non-zero wave vectors $\pm k_0$ (two degenerate ground states). The positions of the dispersion minima according to the effective Hamiltonian (see Eq.~(1) in main text) are given by the following expression:
\begin{equation}
k_0=\sqrt{\frac{\hbar^4\left(2\alpha^2+\gamma\Delta\right)-4M^2\gamma^2\left(2\alpha^2+\gamma\Delta\right)-2\sqrt{\hbar^4\alpha^2\left(\alpha^2+\gamma\Delta\right)\left(\hbar^4-4M^2\gamma^2\right)}}{4M^2\gamma^4-\gamma^2\hbar^4}}
\label{dkveq}
\end{equation}
where $\Delta = E_V-E_H$ is the detuning between the H- and V-polarized modes, which is controlled by the applied voltage. An approximate expression neglecting $\gamma$ is given in the main text (see Eq.~2). We have performed experiments above the condensation threshold, with the formation of a condensate in the two dispersion minima $\pm k_0$. An example of measured time-integrated emission intensity distribution as a function of wave vector $k_x$ and energy is shown in Fig.~4 of the main text, with the intensity plotted as a false color in log scale, in order for the RDSOC dispersion to remain visible as compared to the bright emission of the condensate. We extract the difference between the two minima $\Delta k=2k_0$ as a function of voltage. The results are plotted in Fig 4 of the main text. The agreement of the experiment with the theory is very good, including both the square root dependence expected for the transition between one and two minima, and the asymmetry of the curve due to the parabolic H-V splitting $\gamma$.

\section{Single shot stripes measured in total intensity}

\begin{figure}
    \centering
    \includegraphics[width=0.5\linewidth]{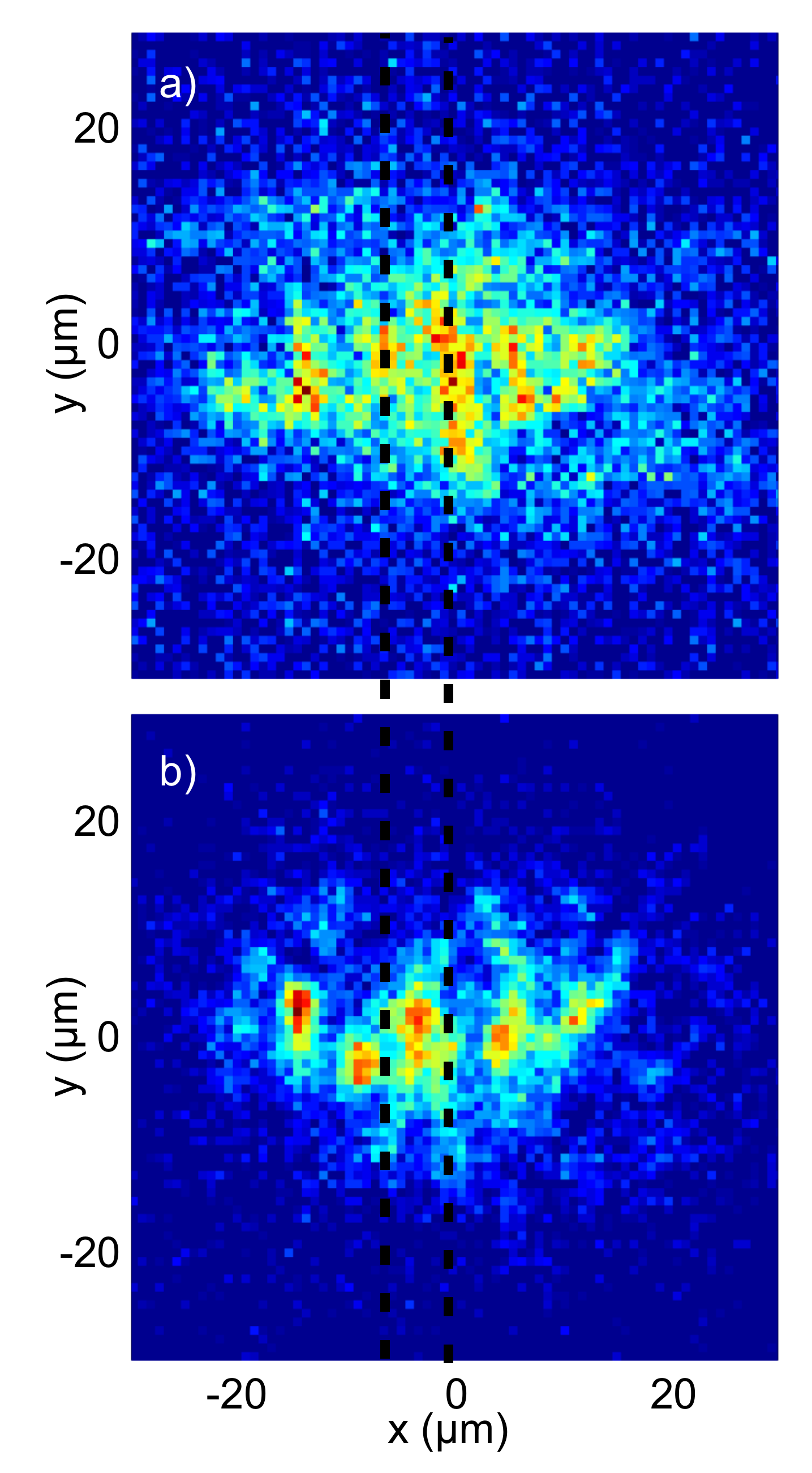}
    \caption{\textbf{Two single-shot images (a,b) showing the total intensity of emission from sample 3 exhibiting stripes.} Black dashed lines are the guides for the eyes allowing to visualize the shift of the stripes due to the spontaneous symmetry breaking. }
    \label{fig:sss3}
\end{figure}

In this section, we present the experimental results obtained on sample "3". These results belong to the set of data used for Fig.~4 of the main text. Figure~\ref{fig:sss3} shows two examples of single-shot images measured in total intensity. They demonstrate the formation of stripes in real space. The positions of these stripes are random, which is highlighted by the black dashed lines allowing to visualize the shift of the stripes between the two panels.

\section{Contrast and statistics}

The mathematical expectation of the contrast of persistent spin helix or density fringes is $C=0$ for averaging over an infinite number of pulses. However, for a finite number of pulses, the contrast is never zero. The law according to which it tends to zero with the increase of the number of realizations can be found as follows. We consider that the intensity profile in each realization is given by $I_i(x)=\cos^2\left(kx+\eta_i\right)$, where $\eta_i$ is a random phase due to spontaneous symmetry breaking. The total intensity for $N$ realizations is given by
\begin{equation}
    I(x)=\sum_{i=1}^N \cos^2\left(kx+\eta_i\right)
\end{equation}
Using trigonometric identities, we transform this to
\begin{equation}
    I(x)=\frac{N}{2}+\frac{1}{2}\left(\cos 2kx \sum_{i=1}^N \cos 2\eta_i - \sin 2kx \sum_{i=1}^N \sin 2\eta_i\right)
\end{equation}
The two sums are correlated: they describe $x$ and $y$ coordinates of a random walker in 2D, which for large $N$ gives $\sum_{i=1}^N \cos 2\eta_i \sim \sqrt{N}\cos 2\theta$, $\sum_{i=1}^N \sin 2\eta_i \sim \sqrt{N}\sin 2\theta$ where $2\theta$ is a random direction from the origin towards the final position of the walker. This allows writing
\begin{equation}
    I(x)=\frac{N}{2}+\frac{\sqrt{N}}{2}\cos(2kx+2\theta)
\end{equation}
which gives the contrast
\begin{equation}
    C(N)=\frac{I_{max}-I_{min}}{I_{max}+I_{min}}=\frac{\sqrt{N}}{N}=\frac{1}{\sqrt{N}}
\end{equation}

\begin{figure}
    \centering
    \includegraphics[width=0.7\linewidth]{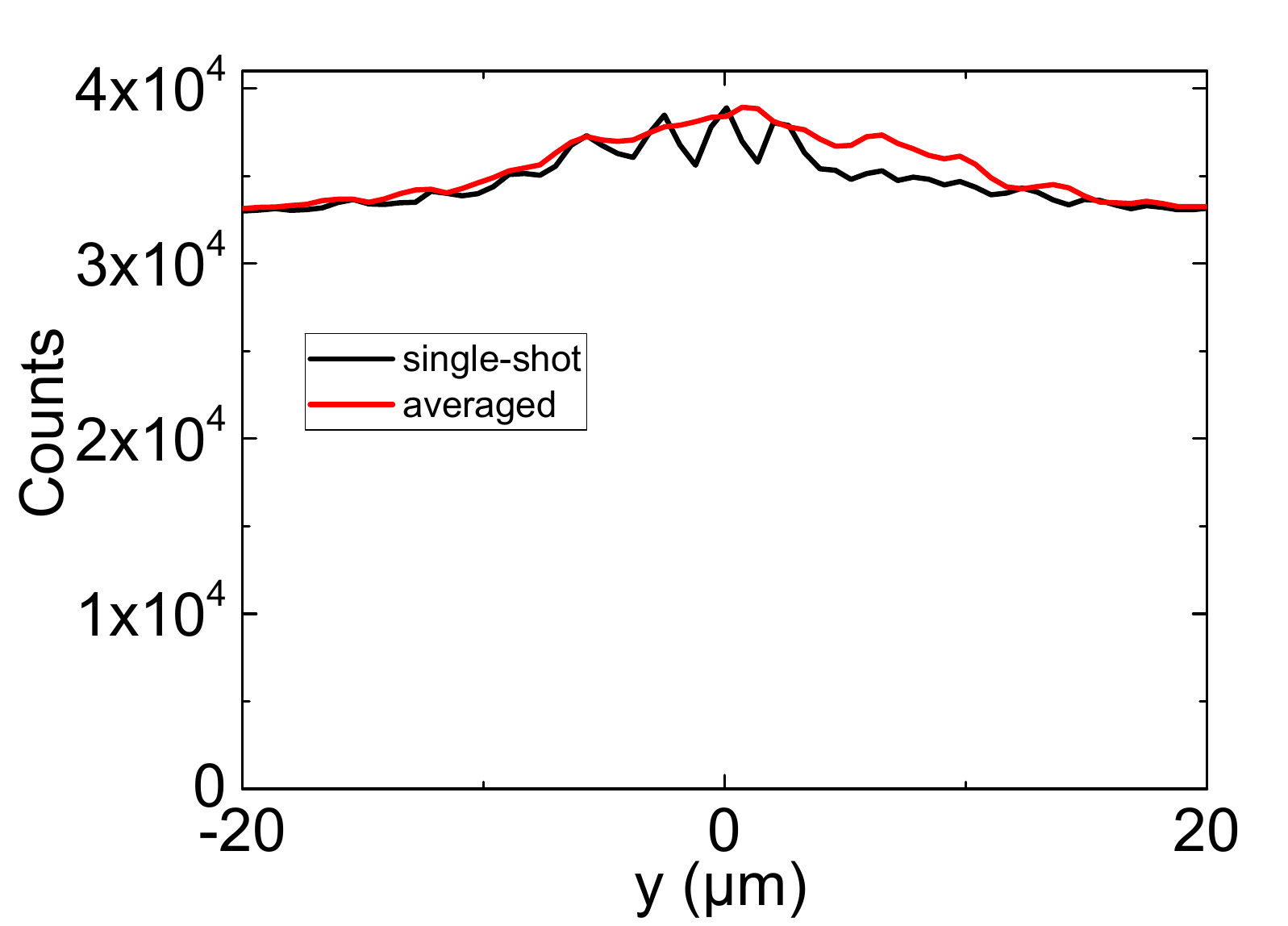}
    \caption{The intensity $I(y)$ for one single-shot measurement (black) and for the averaged case (red), where the contrast is smeared out due to spontaneous symmetry breaking.}
    \label{fig:contrast}
\end{figure}

In Figure~\ref{fig:contrast} we show an example of an intensity profile $I(y)$ for one given single-shot measurement (black) along a single line $x=const$, compared with the result of averaging (red) over all available single-shot images (over 500 for this experiment). No clear stripes are observed in the averaged case.

The decrease of the contrast calculated above can be seen as an application of the Berry-Esseen theorem~\cite{berry1941accuracy,esseen1956moment} to describe how the sample distribution of the random walk tends to the normal distribution. This allows us to extend the consideration to the case of vortex formation via the Kibble-Zurek mechanism, leading to the appearance of dislocations in the interference pattern, which strongly reduces the contrast observed via the k-space intensity distribution, because of the broadening of the peaks.
For random variables with non-identical distributions, the theorem states that the resulting distribution tends towards the normal distribution as $\left(\sum\limits_{i=1}^N\sigma_i^2\right)^{-1/2}$. For identical $\sigma$, this gives the $1/\sqrt{N}$ law obtained above. However, the realizations containing vortices strongly differ from the realizations that do not contain one. The number of vortices in a given realization follows the Poisson distribution: $P_k=\lambda^k e^{-\lambda}/k!$ (the Kibble-Zurek mechanism predicts the average density of vortices for a given set of parameters, and the Poissonian distribution predicts their number in a finite-size system corresponding to this density). We assume that the number of vortices is small: $\lambda\sim 1$, and simplify the consideration by assuming at most 1 vortex in the system.
We can therefore group the terms as
\begin{equation}
    \frac{1}{\sqrt{\sum\limits_{i=1}^N\sigma_i^2}}=\frac{1}{\sqrt{n_0\sigma_0^2+n_1\sigma_1^2}}
\end{equation}
with $n_0+n_1=N$ ($n_0$ is the number of realizations with zero vortices, $n_1$ with 1 vortex). At high $N$, the number of vortices converges to the Poisson distribution, and we can write $n_1=NP_1\approx N/e$, in which case one can factorize by $\sqrt{N}$ and obtain
\begin{equation}
    C\sim\frac{1}{\sqrt{N}\sqrt{\sigma_0^2+e^{-1}\sigma_1^2}}
\end{equation}
meaning that the $N^{-1/2}$ power law is recovered at large $N$ even in presence of vortices. The cases with more than 1 vortex allowed can be treated similarly, as soon as their total number is proportional to $N$ according to the Poisson distribution.

However, for a finite realization set one can expect a deviation from the average number of vortices. This deviation between the set distribution and an ideal probability distribution is itself described by the Berry-Esseen theorem. The number of realizations with a single vortex can therefore be expected to behave as
\begin{equation}
    n_1\sim \left(1+A/\sqrt{N}\right)NP_1
\end{equation}

\begin{figure}
    \centering
    \includegraphics[width=0.7\linewidth]{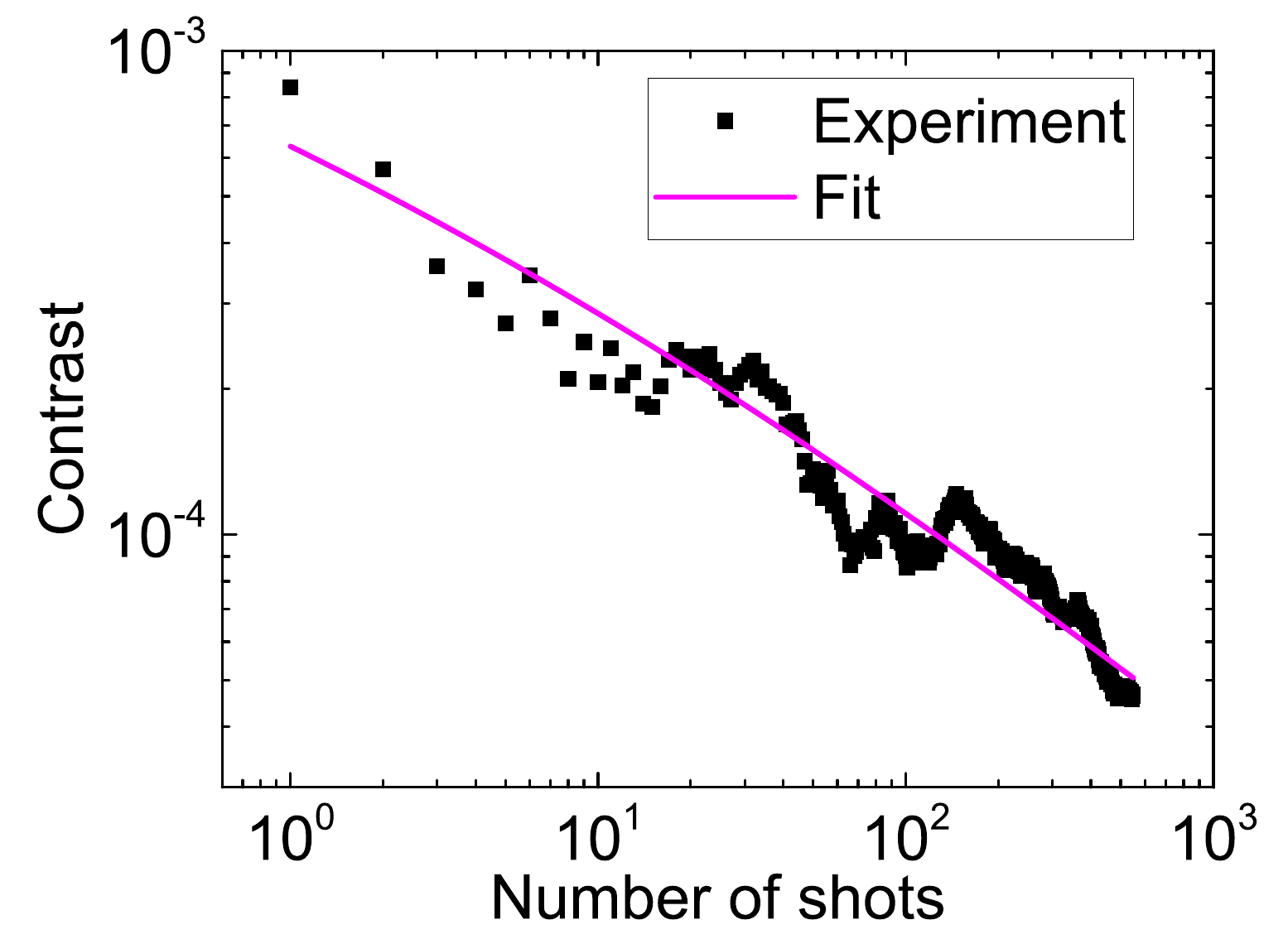}
    \caption{\textbf{The experimentally-measured contrast as a function of number of averaged single shots and its theoretical fit based on Eq.~\eqref{sqfit}.}}
    \label{fig:csq}
\end{figure}

The overall expression for the contrast for a finite number of realizations can be written as
\begin{equation}
    C\sim \frac{A}{\sqrt{N+B\sqrt{N}}}
    \label{sqfit}
\end{equation}
where $A$ and $B$ are fitting parameters. Qualitatively, the power law decrease of the contrast appears slower for small $N$ and tends to $N^{-1/2}$ for large $N$. The transition between the two regimes is determined by $NP_1\sim 1$ (that is, when there is on average at least one vortex in $N$ realizations). This function is used in Fig.~\ref{fig:csq} showing the same data as in Fig.~4 of the main text (sample "2").

We have checked the validity of this model by comparing it with additional large scale numerical simulations (up to $N=10^5$,  with additional averaging) involving periodic stripes with random vortices with a Poissonian distribution.

\section{Interaction mechanisms}

We stress that in order to explain the superfluidity leading to the motion of the stripes, the interactions should be of the order of 1-2~meVs, comparable with the disorder magnitude. Direct exciton-exciton interaction is small in organic materials and also in large band gap semiconductors. For these materials, polariton-polariton interactions are mediated by the screening of the exciton oscillator strength by carrier phase space filling, which provides interaction energies proportional to the Rabi splitting, which is huge for these materials. The magnitude of the effect and its efficiency were recently demonstrated in GaN-based structures, where direct exciton-exciton interactions are certainly weak, but where a polariton mode-locked laser has been reported~\cite{Souissi:24}, clearly demonstrating polariton-polariton interaction in the meV range.

\section{Bogolon dispersion of the supersolid}

\begin{figure}
    \centering
    \includegraphics[width=1.0\linewidth]{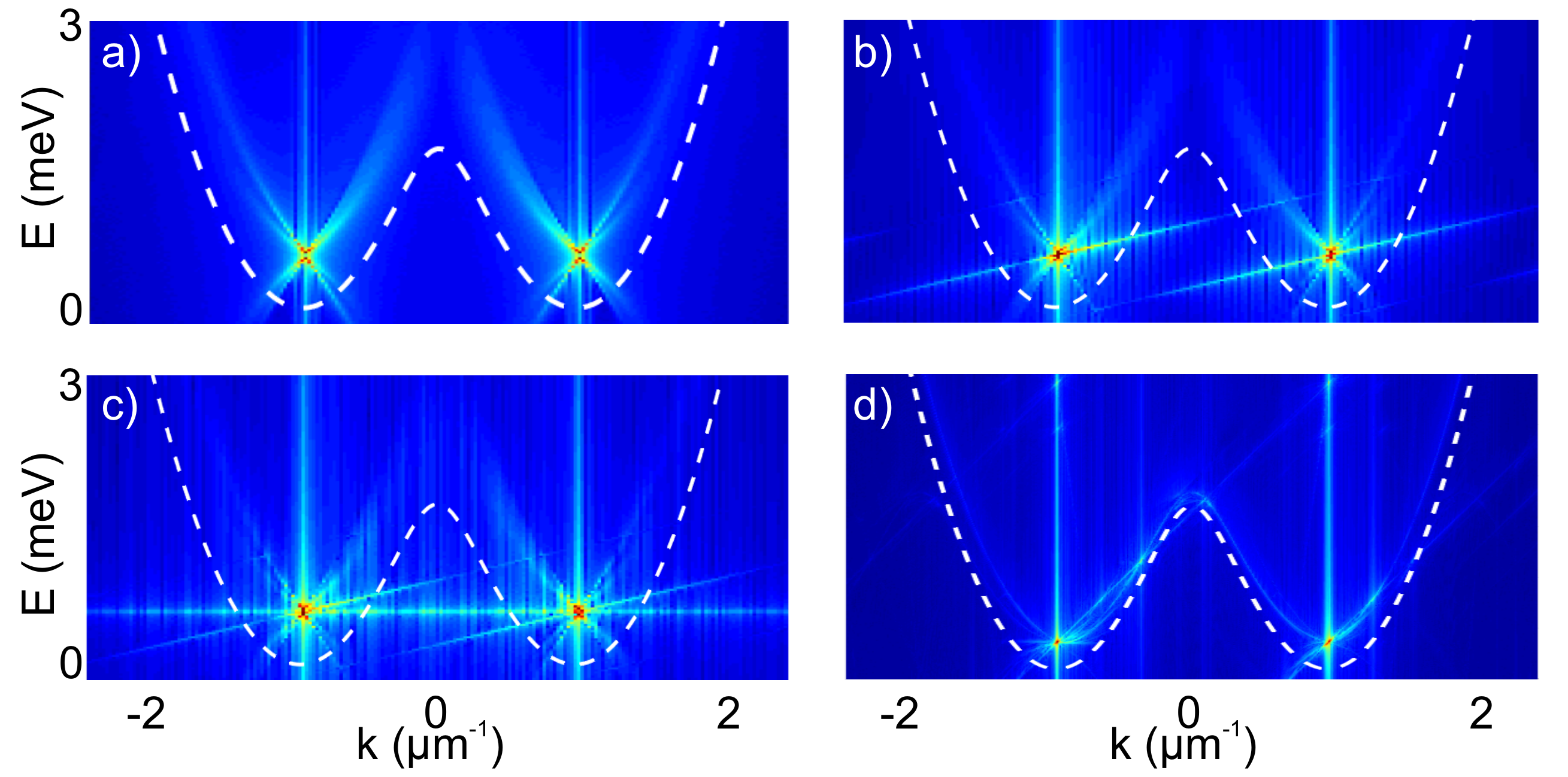}
    \caption{\textbf{Superfluid and supersonic regimes in a supersolid.} Probability density as a function of energy and wave vector $|\psi(k_x,E)|^2$:   a) Bogoliubov dispersion with two different sound velocities visible due to spin-anisotropic interactions; b) subsonic defect propagation at velocity $v<c$, where $c$ is the speed of sound, no resonant scattering from the condensate; c) subsonic defect propagation $v<c$ in presence of a stationary disorder potential, no resonant scattering from the condensate: the condensate is still superfluid;
    d) supersonic defect propagation $v>c$, visible resonant scattering from the condensate. }
    \label{figSSuper}
\end{figure}

We have performed additional numerical simulations in order to demonstrate that the condensate formed in presence of the RDSOC can, at least in principle, exhibit superfluid properties.
We use the same Gross-Pitaevskii equation as in Eq.~(9) from main text (Methods). However, we focus on an idealized case of a homogeneous infinite system that we simulate by using periodic boundary conditions and a homogeneous pump. We also add a small propagating potential defect $U(x,y,t)=U_{def}\exp(-(x-vt-x_0)^2/2\sigma_d^2)\exp(-y^2/2\sigma_d^2)$ instead of the reservoir potential (which we choose as the zero energy reference because it is now homogeneous). Here, $U_{def}$ is the amplitude of the defect potential, $(x_0,0)$ is its initial position in the XY plane, $v$ is its velocity. We plot the Fourier transform of the solution (the probability density as a function of the momentum and energy) in Fig.~\ref{figSSuper}. Panel (a) shows the bogolon dispersion without a moving defect (here we take the interaction constant $g_2=-0.1g_1$ to demonstrate two different sound velocities due to spin-anisotropic interactions). Panel (b) shows the subsonic propagation case. The defect perturbs the density profile of the condensate. This localized perturbation appears in the Fourier transform as a tilted straight line (with the angle corresponding to the velocity of the defect) crossing the condensate. If this line does not cross the bogolon dispersion, the scattering on the defect cannot perturb the condensate and there is no friction or viscosity. This configuration (panel b) corresponds to the Landau criterion of superfluidity in the reference frame of immobile condensate and propagating defect. Panel (c) shows a case with both a moving defect and a static disorder potential, corresponding to the disorder observed experimentally with an optical microscope (Fig.~\ref{figSdisorder}). The superfluid density adjusts itself to the disorder potential profile and the line appearing horizontal is the Fourier transform of the modulated superfluid density.
Panel (d) is with no disorder, but with a defect moving faster than the speed of sound. The tilted straight line corresponds to the velocity of the defect $v$ and crosses the dispersion of bogolons. Resonant (in the frame moving with the defect) scattering from the condensate to the bogolon states can be observed in this case, which corresponds to losses due to friction. Panels (b,c) therefore demonstrate the possibility of superfluid defect motion in a supersolid, as compared with panel (d).

\section{Movies of numerical simulations}
In this section, we present additional Supplementary Movies. 
\begin{itemize}
    \item The movie "vortex.mp4" presents the numerically calculated polariton density as a function of time without disorder potential. A vortex appears during the condensation via the Kibble-Zurek mechanism, it is clearly visible as a dislocation of the stripes pattern in the middle of the figure.
    \item The movie "disorder.mp4" presents the numerically calculated polariton density as a function of time in presence of disorder.
\end{itemize}

\begin{figure}
    \centering
    \includegraphics[width=0.7\linewidth]{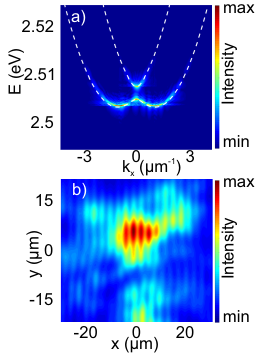}
    \caption{Realistic numerical simulations showing the time-averaged intensity distribution a) as a function of wave vector and energy (the bare polariton dispersion is plotted as a dashed line) and b) in real space, exhibiting clear stripes.}
    \label{fig:thss}
\end{figure}

\begin{figure}[h]
    \centering
    \includegraphics[width=0.95\linewidth]{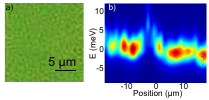}
    \caption{a) An image of a sample with MeLPPP obtained with an optical microscope, demonstrating a visible presence of disorder. b) Spatially-resolved spectrum of polariton modes demonstrating the variation of energy on a few meV scale with micrometer correlation length. }
    \label{figSdisorder}
\end{figure}

We present the results of realistic numerical simulations corresponding to the second movie (the one with a disorder potential) in Fig.~\ref{fig:thss}. These simulations, based on the same Gross-Pitaevskii equation as above (Eq.~(9) from main text) include pulsed pumping and disorder. The two panels of the figure show that the condensate indeed forms in the two minima of the dispersion at $\pm k_0$ and that the time-averaged intensity exhibits stripes, corresponding to the formation of a supersolid. The parameters of these simulations, corresponding to Fig.~3 of the main text, are similar to those of Fig.~4, except the duration of the pulsed excitation, which is short in this case. The time dependencies of the gain and the reservoir potential account for the decay and depletion of the reservoir~\cite{plumhof2014room}.

\end{document}